\documentclass[a4paper, 10pt]{article}
\usepackage{draft}
\usepackage{mathbbol}

\newcommand{\Cocycle}{\gamma}
\newcommand{\hs}{\mathfrak{hs}}
\newcommand{\Invariant}{\mathcal{I}}
\newcommand{\LieBullet}[2]{[#1\,\overset{\bullet}{,}\,#2]_\g}
\newcommand{\LieMetric}[1]{%
    \pmb{\big\langle}#1\pmb{\big\rangle}%
}
\newcommand{\LR}{\underset{\rm LR}{\otimes}}
\newcommand{\BYM}{B}
\newcommand{\BHSYM}{\mathcal{B}}
\newcommand{\CovDerYM}{D}
\newcommand{\CovDerHSYM}{\mathcal{D}}
\newcommand{\FieldStrengthYM}{F}
\newcommand{\FieldStrengthHSYM}{\mathcal{F}}
\newcommand{\GaugeYM}{A}
\newcommand{\GaugeHSYM}{\mathcal{A}}
\newcommand{\LagMult}{\mathtt{Y}}

\usepackage{genyoungtabtikz}
\Yboxdim{8pt}
\Ylinethick{1pt}
\newcommand{\yngSU}[1]{%
    {\Ylinecolour{BurntOrange}\gyoung(#1)}%
}
\newcommand{\yngSO}[1]{%
    {\Ylinecolour{MidnightBlue}\gyoung(#1)}%
}
\newcommand{\yngLab}[2][0pt]{%
  \mathrel{\raisebox{#1}{$\scriptstyle#2$}}%
}
\newcommand{\antiFund}{\Yfillcolour{Peach!60}}

\begin{document}

\title{%
    Massless chiral fields\\
    {\color{MidnightBlue}\adfopenflourishleft}\,
    \raisebox{1.5pt}{%
        \color{PineGreen}$\scriptscriptstyle\blacklozenge$%
    }\,
    {\color{MidnightBlue}\adfopenflourishright} \\
    in six dimensions
}
\abstract{%
    Massless chiral fields of arbitrary spin
    in six spacetime dimensions, also known as higher spin
    singletons, admit a simple formulation
    in terms of $SU^*(4) \cong SL(2,\H)$ tensors.
    We show that, paralleling the four-dimensional case,
    these fields can be described using a $0$-form
    and a gauge $2$-form, taking values in totally symmetric
    tensors of $SU^*(4)$. We then exhibit an example 
    of interacting theory that couples a tower of singletons
    of all integer spin to a background of $\g$-valued
    higher spin fields, for $\g$ an arbitrary Lie algebra
    equipped with an invariant symmetric bilinear form.
    Finally, we discuss the formulation of these models
    in arbitrary even dimensions,
    as well as their partially-massless counterpart.
}
\author{Thomas Basile}
\emailAdd{thomas.basile@umons.ac.be}
\affiliation{%
    Service de Physique de l’Univers, Champs et Gravitation,\\
    Universit\'e de Mons,\\
    20 place du Parc, 7000 Mons, Belgium
}
\maketitle

\section*{Introduction}

Massless fields of arbitrary spin in flat spacetime
can be described in plenty of manners,
meaning with various sets of field variables, at the free level.
Introducing consistent interactions between them 
is, however, notoriously challenging, and the degree
of difficulty of this task may dependent on the set
of field variables chosen for this purpose%
---see e.g. \cite{Bekaert:2010hw, Bekaert:2022poo}
for a broad overview of the problem of constructing
theories with massless higher spin fields,
and \cite{Ponomarev:2022vjb} for an introductory review.

Recently, a \emph{chiral} theory of higher spin fields
in four dimensions has been proposed, first in the light-cone
\cite{Ponomarev:2016lrm} (building upon earlier works
\cite{Metsaev:1991mt, Metsaev:1991nb}),
then in a Lorentz covariant form
\cite{Skvortsov:2022syz,Sharapov:2022faa}.
This theory exhibits an number of interesting features:
For one thing, it admits a formulation around flat space,
a notoriously difficult task to achieve for theories
with massless higher spin fields. Second, it exhibits
signs of good quantum behaviour, such as the absence
of loop corrections and UV divergences in flat space
\cite{Skvortsov:2018jea, Skvortsov:2020wtf, Skvortsov:2020gpn}.
Last, but not least, it shows signs of relation
with `formality-type' theorems
\cite{Sharapov:2022wpz, Sharapov:2022nps, Sharapov:2023erv},
in the following sense: not only does it use 
the Felder--Feigin--Shoikhet $2$-cocycle \cite{Feigin:2005}
whose explicit formula is obtained as a byproduct
of the Tsygan--Shoikhet formality theorem
\cite{Tsygan:1999, Shoikhet:2000gw} (itself a `spin-off'
of Kontsevich's celebrated formality theorem 
\cite{Kontsevich:1997vb}) but it also relies
on an $A_\infty$-algebra whose brackets are expressed
in terms of graphs and weights computed as integrals
over particular configuration spaces,
and whose higher Jacobi identities can be proved
using Stokes' theorem on such spaces---a feature
of the algebraic structure and proofs involved 
in the aforementioned formality theorems.

Chiral higher spin gravity admits two contractions
which are higher spin extensions of self-dual Yang--Mills
and self-dual gravity
\cite{Ponomarev:2017nrr, Krasnov:2021nsq, Krasnov:2021cva, Skvortsov:2022unu},
and hence displays (strong) ties to twistor theory 
\cite{Tran:2021ukl, Tran:2022tft, Herfray:2022prf, Adamo:2022lah, Tran:2022mlu, Steinacker:2022jjv, Neiman:2024vit}. The latter relates a complexified
(and compactified) version of spacetime in four dimensions
to another complex manifold referred to as `twistor space'
(see \cite{Penrose:1967wn, Penrose:1968me, Penrose:1972ia, Hughston:1979tq, Penrose:1985bww, Penrose:1986ca, Ward:1990vs, Adamo:2017qyl} and references therein).
This \emph{twistor correspondence} relies on the fact
that both manifolds are homogeneous spaces of $SL(4,\C)$,
which contains the double cover of the Lorentz group
\parbox{90pt}{$\widetilde{SO}(1,3) \cong SL(2,\C)$}. 
This allows for a simple description of massless fields%
\footnote{Note that massive fields can also be described
in terms of tensor-spinors,
see e.g. \cite{Hughston:1979tq, Ochirov:2022nqz, Delplanque:2024enh}.}
in terms of symmetric tensor-spinors
(e.g. \cite{Eastwood:1981jy, Eastwood:1987ki}),
at the basis of chiral higher spin gravity.

A similar isomorphism holds in six dimensions:
the (double cover of the) Lorentz group is isomorphic
to the special linear group of $2 \times 2$ matrices
over the quaternions, 
\[
    \widetilde{SO}(1,5) \cong SL(2,\H)\,,
\]
the latter being sometimes presented as the \emph{real form} 
of the special linear group of $4 \times 4$ matrices, $SU^*(4)$. 
The fundamental and anti-fundamental
representation of $SU^*(4)$, which are of dimension $4$,
correspond to the two chiral spinors of the Lorentz group. 
More generally, this isomorphism relates totally symmetric 
$SU^*(4) \cong SL(2,\H)$ tensors (of even ranks)
to irreducible representations (irreps) of Lorentz algebra
characterised by a three-row Young diagram,
all of which being of the same length, thereby making up
a rectangle, and required to be self-dual in each column.
Such tensors, subject to first order differential equations,
can be identified with the self-dual part 
of the curvature of a gauge field with the symmetry
of a two-row rectangular Young diagram.
This particular class of mixed-symmetry fields%
\footnote{Generally, the term \emph{mixed-symmetry fields}
refers to those fields in flat space that fall into
a representation of the Poincar\'e group induced from an irrep
of the massless little group labelled by a Young diagram
with more than one row and one column (see \cite{Bekaert:2006py} for a review
of the representation theory of the Poincar\'e group,
and \cite{Labastida:1986gy, Labastida:1986ft, Labastida:1986zb, Labastida:1987kw, Bekaert:2002dt, Bekaert:2003zq, Bekaert:2006ix, Alkalaev:2008gi, Boulanger:2008up, Boulanger:2008kw, Alkalaev:2009vm, Campoleoni:2008jq, Campoleoni:2009gs, Campoleoni:2009hj, Campoleoni:2009je, Skvortsov:2008vs, Skvortsov:2008sh, Skvortsov:2009nv, Skvortsov:2009zu, Alkalaev:2011zv, Campoleoni:2012th, Bekaert:2015fwa, Bekaert:2023uve}
for the description of massless mixed-symmetry fields,
in both flat and (anti-)de Sitter spacetimes).}
is known as `singletons'
\cite{Angelopoulos:1998, Laoues:1998ik, Angelopoulos:1999bz}
and corresponds exactly to those massless fields
in flat space which are also \emph{conformal}
\cite{Siegel:1988gd, Barnich:2015tma}
(see \cite{Bekaert:2011js} for a pedagogical review
of the peculiar features of singletons).

The fact that these fields are of \emph{mixed-symmetry} type
constitutes one of the interesting differences
with respect to the four-dimensional case.
Finding interacting theories for mixed-symmetry fields
is also challenging, owing to the increase
of gauge symmetries to preserve (in flat space),
and to their reducible character, see e.g.
\cite{Bekaert:2004dz, Boulanger:2011se, Boulanger:2011qt, Joung:2016naf}. They however play a central role
in duality-symmetric formulation of gauge theories
(see for instance \cite{Curtright:1980, Boulanger:2003vs, deMedeiros:2002qpr, Henneaux:2019zod, Boulanger:2020yib}
and references therein).
More specific to six dimensions is the putative existence
of the $(2,0)$ superconformal field theory
\cite{Witten:1995em, Strominger:1995ac, Witten:1995zh, Seiberg:1996vs}
whose spectrum consists of a supermultiplet including
a self-dual $2$-form (see also 
\cite{Samtleben:2011fj, Samtleben:2011drz, Samtleben:2012fb, Palmer:2013pka, Bandos:2013jva, Lavau:2014iva, Saemann:2017rjm, Samann:2017sxo} 
for recent work on the related $(1,0)$ model),
or a singleton of spin $1$ in our terminology;
and the `exotic' supergravity known as the $(4,0)$ theory
\cite{Hull:2000rr, Hull:2000zn, Hull:2001iu}
based on a supermultiplet containing a singleton of spin $2$,
thought of as a graviton, as well as several singletons
of spin $1$ (see for instance
\cite{Henneaux:2016opm, Henneaux:2017xsb, Minasian:2020vxn, Cederwall:2020dui, Bertrand:2020nob, Bertrand:2022pyi}
for recent work on this topic).

Having a `twistor-inspired' description of singletons
at hand begs the question of whether one can reproduce
the success of $4d$ chiral higher spin gravity in $6d$?
In this paper, we take a first step towards addressing
this question by exhibiting a simple formulation
for massless higher spin fields in $6d$ dimensions,
in terms of field variables which takes advantage 
of the isomorphism $\so(1,5) \cong \sl(2,\H) \cong \su^*(4)$, 
and present a couple of examples of interacting theories
involving said massless fields. 

This paper is organised as follows:
in Section \ref{sec:6d} we start by reviewing
the free description of higher spin singletons
in $6d$ and show that they admit a formulation
based on a pair of $0$-form and a $2$-form valued
in symmetric $\su^*(4)$-tensors, then use this formulation 
as a starting point to couple higher spin singletons
to a connection $1$-form, and define an higher spin extension 
thereof. In Section \ref{sec:any_d}, we discuss how to
extend the free formulation discussed in $6d$
to arbitrary even dimensions, and argue that
the interacting theories also admit
a higher-dimensional counterpart.
We conclude this paper in Section \ref{sec:discu}
by discussing various future directions,
and propose a `partially-massless' version of the theory 
discussed in the bulk of this note in Appendix \ref{app:PM}.

\paragraph{Conventions.}
In order to lighten expressions containing tensors
of arbitrary ranks, we use a notational shorthand
common in the higher spin literature which consists in
denoting all indices to be symmetrised by the same letter,
and indicating the number of such indices in parenthesis
if necessary. For instance, we write the components
of a symmetric tensor $T$ of rank $n$ as
\[
    T_{a(n)} = \tfrac1{n!}\,\sum_{\sigma\in\cS_n}
    T_{a_{\sigma_1} \dots a_{\sigma_n}}\,,
\]
where $\cS_n$ is the group of permutations of $n$ object.
We also write all tensors with the symmetry
a given Young diagram in the \emph{symmetric basis}.
In plain words, when saying that a tensor has the symmetry
of a diagram with $k$ rows of length
$\ell_1 \geq \ell_2 \geq \dots \geq \ell_k > 0$,
we mean that this tensor has $k$ groups of symmetric indices,
the first one with $\ell_1$ indices, the second with $\ell_2$,
etc., and that the symmetrisation of all indices
in the $i$th group with one index of the $j$th group
with $j>i$ vanishes identically. Such a tensor
is therefore denoted by
\[
    T_{a_1(\ell_1),\dots,a_k(\ell_k)}
    \with
    T_{\dots,a_i(\ell_i),\dots,a_i\,a_j(\ell_j-1),\dots}=0\,,
\]
using the previously introduced notation.
In particular, different groups of indices
separated by a comma obey the over-symmetrisation condition
given above, while groups of indices separated by
a vertical bar do not obey any symmetrisation condition
(i.e. $T_{a(\ell)|b}$ \emph{does not vanish}
when symmetrising all indices, while $T_{a(\ell),b}$ does).
In particular, this means that we denote a pair
of antisymmetric indices by separating them with a comma,
$T_{a,b}=-T_{b,a}$.

\section{Six dimensions}
\label{sec:6d}
\subsection{Free higher spin singletons}
\label{sec:free}
As the chiral formulation of massless higher spin field
in $4d$ presented in e.g. \cite{Krasnov:2021nsq} relies
on the isomorphism of Lie algebras $\so(1,3) \cong \sl(2,\C)$,
its $6d$ counterpart that we will present hereafter
is based on the isomorphism $\so(1,5) \cong \sl(2,\H)$,
the latter algebra being also isomorphic to $\su^*(4)$.
Since finite-dimensional representations of both algebras
can be conveniently denoted by Young diagrams,
we will use the convention that diagrams
corresponding to $\so(1,5)$ irreps will be drawn
in \textcolor{MidnightBlue}{blue}, while diagrams
corresponding to $\su^*(4) \cong \sl(2,\H)$-irreps
will be drawn in \textcolor{Peach}{orange}.
For instance, we will write the isomorphism between
the vector irrep of $\so(1,5)$ with the antisymmetric
rank $2$ tensor of $\su^*(4)$ as
\begin{equation}
    \underset{\so(1,5)}{\yngSO{;}}
    \quad \cong \quad
    \underset{\su^*(4)}{\yngSU{;,;}}\,,
\end{equation}
and will forget about the subscripts $\so(1,5)$
and $\su^*(4)$ from now on.

The above isomorphism tells us that, similarly
to the $d=4$ case, we can use coordinates carrying
pairs of antisymmetric spinor indices
$A, B, \dots = 1,\dots,4$, i.e.
\begin{equation}
    x^\mu \leadsto x^{A,B} = -x^{B,A}\,,
\end{equation}
and the same holds for vector fields and differential forms.
Indices can only be raised and lowered by pairs, 
using the Levi--Civita symbol $\varepsilon_{ABCD}$
of $\su^*(4)$, via
\begin{equation}\label{eq:indicesA3}
    v_{A,B} := \tfrac12\,\varepsilon_{ABCD}\,v^{C,D}
    \qquad\Longleftrightarrow\qquad
    v^{A,B} = \tfrac12\,\varepsilon^{ABCD}\,v_{C,D}\,.
\end{equation}
This invariant tensor also allows us to reduce any diagram
to one with \emph{at most two rows}. Indeed, we can convert
any column with three boxes into a single box, via
\begin{equation}\label{eq:dualisation}
    \yngSU{;,;,;} \ni v_{A,B,C}
    \quad\longmapsto\quad
    \tilde v^A\ \propto\ \varepsilon^{ABCD}\,v_{B,C,D}\,
    \in \yngSU{!\antiFund;}\,,
\end{equation}
where we filled in the box to keep in mind that
the corresponding tensor has an index up:
it belongs to the $\su^*(4)$-irrep \emph{conjugate} 
to the one denoted by a box. We will adopt this notation 
in the reminder of this note, namely uncoloured boxes 
will correspond to tensors of the vector representation, 
and coloured boxes to tensors of its conjugate.
For the antisymmetric Lorentz tensor of rank $2$,
the relevant isomorphism is
\begin{equation}\label{eq:adjoint}
    \yngSO{;,;} \quad \cong \quad \yngSU{;;,;,;}
    \quad \cong \quad \yngSU{:;,!\antiFund;}
\end{equation}
where in the last step, we have dualised the first column
as in \eqref{eq:dualisation}. Note that the corresponding
tensor is traceless, in the sense that the contraction
between the upper and lower indices vanishes.
One can recognize the adjoint representation of $\so(1,5)$
and $\su^*(4)$ in the above isomorphism: one way to present
the Lorentz algebra is as being generated by $M_{a,b}=-M_{b,a}$
with $a,b=0,1,\dots,5$ subject to the relations
\begin{equation}
    [M_{a,b}, M_{c,d}]=\eta_{bc}\,M_{ad} - \eta_{bd}\,M_{a,c}
    - \eta_{ac}\,M_{b,d} + \eta_{ad}\,M_{b,c}\,,
\end{equation}
whereas for $\su^*(4)$, one can use generators $M^A{}_B$
with $A,B=1,\dots,4$ such that $\delta_A^B\,M^A{}_B=0$,
and subject to the relations
\begin{equation}
    [M^A{}_B, M^C{}_D]
    = \delta_B^C\,M^A{}_D - \delta_D^A\,M^C{}_B\,.
\end{equation}
Recall that the finite-dimensional \emph{irreducible}
representations of special linear algebras correspond 
to tensors whose components may have both upper 
and lower indices, both groups having the symmetry
of a given (generally different) Young diagram, 
and which are \emph{traceless} in that the contraction
of any upper index with any lower one vanishes identically.
From now on, we will denote these irreps
by drawing the two diagrams, one with its boxes
filled in and the other not, and joining the upper right corner
of the former one with the lower left corner of the latter%
---as we did for the adjoint representation \eqref{eq:adjoint}.

Finally, let us note that for $\so(1,5)$, 
totally antisymmetric tensors of rank $3$
can be either self-dual or anti-self-dual,
which are isomorphic to two different irreps
of $\su^*(4)$, namely
\begin{equation}
    \yngSO{;,;,;}_+ \quad\cong\quad \yngSU{;;}
    \qand
    \yngSO{;,;,;}_- \quad\cong\quad \yngSU{;;,;;,;;}
    \quad\cong\quad \yngSU{!\antiFund;;}
\end{equation}
where in the second term we again dualised 
the two columns of height $3$ to get a symmetric 
tensor of rank $2$.

Assume that the six-dimensional manifold $M$
on which we are working is equipped with a vielbein $e^{A,B}$,
which is a $1$-form taking values in the antisymmetric irrep
of $\su^*(4)$. Following \eqref{eq:indicesA3}, the vielbein
with all indices lowered is defined as, 
\begin{equation}
    e_{A,B} := \tfrac12\,\varepsilon_{ABCD}\,e^{C,D}\,,
\end{equation}
and we can use it to define a basis for the space
of $2$-forms, via
\begin{equation}
    \Sigma_A{}^B := e_{A,C} \wedge e^{C,B}
    \equiv \tfrac12\,\varepsilon_{ACDE}\,
    e^{B,C} \wedge e^{D,E}\,.
\end{equation}
In particular, one can easily see that it is traceless,
so that these $2$-forms do carry the irrep
$\Yboxdim{5pt}\yngSU{:;,!\antiFund;}$\,,
and a direct computation leads to
\begin{equation}\label{eq:product}
    e^{A,B} \wedge e^{C,D}\ = \tfrac12\, 
    \big(\varepsilon^{\times AB[C}\,\Sigma_\times{}^{D]}
    - \varepsilon^{\times CD[A}\,\Sigma_\times{}^{B]}\big)
    \qquad\Longleftrightarrow\qquad 
    e^{A,B} \wedge e_{C,D}
    = 2\,\delta_{[C}^{[A}\,\Sigma_{D]}{}^{B]}\,,
\end{equation}
thereby allowing us to write any product of $1$-forms
in this basis. We can proceed further and define
the following anti/self-dual $3$-forms,
\begin{subequations}
    \label{eq:3-forms}
    \begin{align}
        H^{AA} & := e^{A,C} \wedge \Sigma_C{}^A
        \equiv e_{B,C} \wedge e^{A,B} \wedge e^{A,C}\,,\\
        H_{AA} & := \Sigma_A{}^C \wedge e_{C,A}
        \equiv e^{B,C} \wedge e_{A,B} \wedge e_{A,C}\,,
    \end{align}
\end{subequations}
which form a basis of the space of $3$-forms.
Once again, a direct computation yields
\begin{equation}
    e^{A,B} \wedge \Sigma_D{}^C
    = \tfrac13\,\varepsilon^{ABC\times}\,H_{\times D}
    -\tfrac23\,\delta^{[A}_D\,H^{B]C}_{\,}\,,
\end{equation}
thereby expressing the product of any $1$-form
with any other $2$-form in the basis of $3$-forms
defined in \eqref{eq:3-forms}. Next, we can define
a basis of $4$-forms by taking products
of the previously obtained basis of $1$-, $2$- and $3$- forms,
projected on the correct $\su^*(4)$-irrep
(that is $\Yboxdim{5pt}\yngSU{:;,!\antiFund;}$), namely
\begin{equation}
    H_{AC} \wedge e^{C,B}
    = e_{A,C} \wedge H^{CB}
    = \Sigma_A{}^C \wedge \Sigma_C{}^B\,.
\end{equation}
Yet again, a direct computation gives us
\begin{equation}\label{eq:1times3form}
    e^{A,B} \wedge H_{CC}
    = 2\,\Sigma_C{}^D \wedge \Sigma_D{}^{[A}\,\delta^{B]}_C\,,
    \qquad\text{and}\qquad
    e_{A,B} \wedge H_{CC} = 
    \Sigma_C{}^D \wedge \Sigma_D{}^\times\,
    \varepsilon_{\times ABC}\,,
\end{equation}
so that, combined with the previous results, 
we can decompose products of forms up to degree $4$
in the basis introduced so far.
In particular, one can check that
\begin{equation}\label{eq:3times2form}
    H_{AA} \wedge e_{A,B} = 0
    \qquad\Longrightarrow\qquad 
    H_{AA} \wedge \Sigma_A{}^B = 0\,,
\end{equation}
as a consequence of \eqref{eq:product}.
Let us conclude this discussion on forms
by pointing out that the self-dual $3$-forms
are `orthogonal' to one another, whereas
the product of self-dual and anti-self-dual $3$-forms
is proportional to a top form on $M$,
\begin{equation}\label{eq:product3forms}
    H_{AA} \wedge H_{BB} = 0\,,
    \qquad 
    H_{AA} \wedge H^{BB}
    \propto \delta_A^B\,\delta_A^B\,{\rm vol}_M\,,
\end{equation}
where ${\rm vol}_M$ denotes the volume form
induced by the vielbein.

\paragraph{Massless equations in six dimensions.}
Around flat spacetime in $d=6$ dimensions,
there are several types of massless fields
one can consider, since their spin is determined by
an irrep of $Spin(4) \cong SU(2) \times SU(2)$,
and hence labelled by \emph{two (half-)integers}.
This opens the possibility of considering
\emph{mixed-symmetry} fields, which are fields
whose little group irrep corresponds
to a two-row Young diagram of $Spin(4)$, or equivalently,
for which the two $SU(2)$ irreps are different.
The `limiting case' when exactly one of this $SU(2)$
irreps is trivial, corresponding to \emph{rectangular}
two-row Young diagrams.

Such fields can be realised by Lorentz tensors,
with the symmetry of the same two-row diagram,
and enjoying some gauge symmetries. The simplest example
is probably a massless $2$-form $B$, whose curvature 
$3$-form is denoted $H$,
\begin{eqnarray}
    B = B_{\mu,\nu}\, {\rm d}x^\mu \wedge {\rm d}x^\nu
    \in \Omega^2_M \cong \yngSO{;,;}
    \hspace{75pt}
    H = {\rm d}B \in \Omega^3_M \cong \yngSO{;,;,;}\,,
\end{eqnarray}
on which we can impose some self-duality
or anti-self-duality condition. More generally, 
one can consider the spin $s$ version, meaning 
a gauge field $\varphi$ with the symmetry
of a rectangular two-row diagram of length $s$,
\begin{align}
    \varphi_{\mu(s),\nu(s)}\ \longleftrightarrow\ \yngSO{_5<\yngLab[1pt]{s}>,_5}\ \cong\
    \yngSU{_5_5<\yngLab[1pt]{s}>,_5<\yngLab[1pt]{s}>,_5}\
    \cong\ \yngSU{:::::_5<\yngLab[1pt]{s}>,%
    !\antiFund_5<\yngLab[1pt]{s}>}\
    \longleftrightarrow\ \varphi_{A(s)}{}^{B(s)} 
\end{align}
whose curvature is a rectangular three-row diagram,
\begin{align}\label{eq:field-strength}
    F_{\mu(s), \nu(s), \rho(s)}
    = \underbrace{\partial_\rho \dots \partial_\rho}_{s}
    \varphi_{\mu(s), \nu(s)} + \dots & \longleftrightarrow\
    \yngSO{_5,_5<\yngLab[1pt]{s}>,_5} \\ & \cong\
    \yngSU{_6<\yngLab[1pt]{2s}>}\ \longleftrightarrow\ F_{A(2s)}
    = \underbrace{\partial_{A,B} \dots \partial_{A,B}}_{s}
    \varphi_{A(s)}{}^{B(s)}\,, \nonumber \\ & \oplus\
    \yngSU{!\antiFund_6<\yngLab[1pt]{2s}>}\ \longleftrightarrow\ F^{B(2s)}
    = \underbrace{\partial^{A,B} \dots \partial^{A,B}}_{s}
    \varphi_{A(s)}{}^{B(s)}\,, \nonumber
\end{align}
on which we can also impose anti/self-duality conditions. 
In terms of $\su^*(4)$, this becomes quite simple
as it amounts to keeping either $F_{A(2s)}$ or $F^{A(2s)}$ only.

The aforementioned massless fields are known to be conformal,
and are sometimes referred to as spin $s$ singletons 
\cite{Angelopoulos:1998, Angelopoulos:1999bz}.%
\footnote{Note that they were already identified
in \cite{Siegel:1988gd}, and discussed in more details
recently \cite{Bandos:2005mb, Bekaert:2009fg, Bekaert:2011js}.
See also \cite{Gunaydin:1999ci, Sezgin:2001ij, Bekaert:2009fg, Fernando:2010dp, Fernando:2010ia, Govil:2014uwa}
for more details concerning their symmetry algebra,
and \cite{Joung:2014qya, Campoleoni:2021blr}
for reviews of higher spin algebras in general dimensions.}
The equations of motion for a singleton of spin $s$
and say positive chirality, in terms of $\su^*(4)$ tensors
read \cite{Bandos:2005mb, Mason:2011nw, Saemann:2011nb}%
\footnote{See also \cite{Bekaert:2009fg} for the formulation
of higher spin singletons in \emph{arbitrary even} dimensions
using Howe duality.}
\begin{equation}\label{eq:EOM_Weyl}
    \partial_{B,C} \Psi^{A(2s-1)C} \approx 0
\end{equation}
or in terms of a potential,%
\footnote{The two are typically related by $\Psi^{A(2s)}
= (\partial^{A,B})^{2s-1}\Phi_{B(2s-1)}{}^A$ in flat space,
with $\Psi^{A(2s)}$ corresponding to a chiral half
of the field strength in \eqref{eq:field-strength}.}
\begin{equation}\label{eq:EOM_potential}
    \partial_{B,A} \Phi_{A(2s-1)}{}^B \approx 0\,,
\end{equation}
where $\Phi_{A(2s-1)}{}^B$ is traceless, and we used
the symbol `\,$\approx$\,' merely as a way to highlight the fact
that solutions of these equations are \emph{on-shell} fields.
The last equation is invariant under the gauge symmetry
\begin{equation}\label{eq:gaugePhi}
    \delta_\xi\Phi_{A(2s-1)}{}^B
    = \partial^{B,C} \xi_{A(2s-1),C}
    - \tfrac{2s-1}{2s+2}\,\delta^B_A \,
    \partial^{C,D}\xi_{A(2s-2)C,D}\,,
\end{equation}
where $\xi_{A(2s-1),B}$ is of `hook-type' symmetry,
$\Yboxdim{5pt}\yngSU{;_3,;}$ (one should use the identity
$\partial_{A,B} \partial^{B,C} \propto \delta^C_A \Box$
and the symmetry property of $\xi$ to verify this statement).
This gauge symmetry is reducible, as one can check that
gauge parameters of the form
\begin{equation}
    \mathring{\xi}_{A(2s-1),B}
        := \partial_{B,A} \zeta_{A(2s-2)}\,,
\end{equation}
lead to trivial gauge transformations,
i.e. $\delta_{\mathring{\xi}}\,\Phi_{A(2s-1)}{}^B = 0$\,.
These wave equations, \eqref{eq:EOM_Weyl}
and \eqref{eq:EOM_potential},
can be derived from the first order action
\begin{equation}\label{eq:actionPhi}
    S[\Psi,\Phi] = \int_M {\rm vol}_M\
    \Psi^{A(2s)}\,\partial_{B,A} \Phi_{A(2s-1)}{}^B\,\,,
\end{equation}
where ${\rm vol}_M$ denotes the volume form
on $d=6$ Minkowski spacetime.

Let us point out that, contrary to the $4d$ case,
the fields $\Psi$ and $\Phi$ both describe a singleton
of spin $s$ with the \emph{same chirality}
(we will comment further on this point
in Section \ref{sec:any_d}), and hence the above functional
and its reformulation that will be presented below
are not actions for self-dual $2$-forms when $s=1$.
Actions for self-dual $p$-forms have been studied
by many authors, see for instance 
\cite{Siegel:1983es, Henneaux:1988gg, Pasti:1996vs, Bekaert:1998yp, Bekaert:1999dp, Bekaert:2000qx, Ho:2012nt, Sen:2015nph, Mkrtchyan:2019opf, Avetisyan:2021heg, Avetisyan:2022zza, Evnin:2022kqn, Evnin:2023ypu}
and references therein.

\paragraph{Action principle for singletons.}
We can re-write a little bit the previous action
by embedding the field $\Phi_{A(2s-1)}{}^B$
in a $2$-form, via
\begin{equation}
    \varpi_{A(2s-2)} = \Sigma_C{}^B\,\Phi_{A(2s-2)B}{}^C\,,
\end{equation}
and assuming that we are now working
on a constant curvature background equipped with
a torsion-free connection, i.e.%
\footnote{We implicitly normalised this curvature
for simplicity.}
\begin{equation}
    \nabla^2 = \Sigma_A{}^B\,\rho(M^A{}_B)\,,
\end{equation}
where $M^A{}_B$ are the $\su^*(4)$-generators
and $\rho$ the representation of this algebra
on the section of the bundle acted upon by the curvature.
Concretely, it verifies
\begin{equation}
    \begin{aligned}
        \nabla^2\varphi_{A(k),B(l)}{}^{C(m),D(n)}
        & = -k\,\Sigma_A{}^\times\,
        \varphi_{A(k-1)\times,B(l)}{}^{C(m),D(n)}
        - l\,\Sigma_B{}^\times\,
        \varphi_{A(k),B(l-1)\times}{}^{C(m),D(n)} \\
        &\ + m\,\Sigma_\times{}^C\,
        \varphi_{A(k),B(l)}{}^{C(m-1)\times,D(n)}
        + n\,\Sigma_\times{}^D\,
        \varphi_{A(k),B(l)}{}^{C(m),D(n-1)\times}\,,
    \end{aligned}
\end{equation}
when acting on a generic tensor $\varphi$
which is `hook-symmetric' in both its upper and lower indices. 
The covariant differential of the $2$-form $\varpi$, 
which is a $3$-form and hence contain both a self-dual
and an anti-self-dual part, reads
\begin{equation}
    \nabla\varpi_{A(2s-2)} 
    = \tfrac23\,\nabla^{B,C}\Phi_{A(2s-2)B}{}^C\,H_{CC}
    + \tfrac23\,H^{BB}\,\nabla_{B,C}\Phi_{A(2s-2)B}{}^C\,,
\end{equation}
as can be checked using \eqref{eq:product}.
Using the identity \eqref{eq:product3forms},
we can get rid of the first term and keep only
the second one in previous equation by multiplying
it by the self-dual $3$-form, 
\begin{equation}\label{eq:potential_form}
    H_{AA} \wedge \nabla\varpi_{A(2s-2)}
    \propto {\rm vol}_M\,\nabla_{B,A} \Phi_{A(2s-1)}{}^B\,,
\end{equation}
thereby reproducing the part of the integrand
of the action \eqref{eq:actionPhi} to be contracted with $\Psi$.

Having re-expressed the action for a spin-$s$ singleton
in terms of differential forms, by embedding $\Phi_{A(2s-1)}{}^B$
in the $2$-form $\varpi_{A(2s-2)}$,
we can now promote the latter to an independent field%
---that we shall denote by $\omega_{A(2s-2)}$
so as to avoid any confusion---and postulate the free action
\begin{equation}\label{eq:free_action}
    S[\Psi,\omega] = \int \Psi^{A(2s)}\,
    H_{AA} \wedge \nabla \omega_{A(2s-2)}\,,
\end{equation}
which looks exactly like the one proposed
in \cite{Krasnov:2021nsq}, upon replacing
$SL(2,\C)$ spinor indices for $SU^*(4) \cong SL(2,\H)$ ones.
This action is invariant under the gauge symmetries
\begin{equation}\label{eq:gauge_transfo}
    \delta_{\xi,\eta} \omega_{A(2s-2)}
    = \nabla \xi_{A(2s-2)} + e_{A,B}\,\eta_{A(2s-3)}{}^B
\end{equation}
as can be verified using the identities \eqref{eq:3times2form}.
Note that, as before, these gauge transformations are reducible:
gauge parameters of the form%
\footnote{Note that in flat space, the reducibility parameter
for $\eta$ becomes trivial, i.e. $\mathring\eta=0$.
Indeed, the role of $\mathring\eta$ is to compensate
for the curvature term appearing due to $\mathring\xi$.
This limit is not explicit here due our choice of absorbing
the cosmological constant in the definition of $H_A{}^B$.}
\begin{equation}\label{eq:reducibility}
    \mathring\xi_{A(2s-2)} = \nabla \zeta_{A(2s-2)}
    \qand
    \mathring\eta_{A(2s-3)}{}^B
    = (2s-2)\,e^{B,C}\,\zeta_{A(2s-3)C}\,,
\end{equation}
leave $\omega$ invariant,
$\delta_{\mathring\xi,\mathring\eta}\omega=0$,
for any $0$-form $\zeta_{A(2s-2)}$. These gauge symmetries
allows one to gauge away the unwanted components
contained in $\omega_{A(2s-2)}
= \Sigma_C{}^B\,\omega_{A(2s-2)|B}{}^C$. To be more precise,
the decomposition into irreducible components
of this $2$-form reads
\begin{equation}
    \yngSU{:;,!\antiFund;} \otimes \yngSU{_5<\yngLab[1pt]{2s-2}>}\,
    \quad\cong\quad \yngSU{:_6<\yngLab[1pt]{2s-1}>,!\antiFund;}\
    \oplus\ \yngSU{:_5<\yngLab[1pt]{2s-2}>,:;,!\antiFund;}\
    \oplus\ \yngSU{_5<\yngLab[1pt]{2s-2}>}\,
    \oplus\ \yngSU{_4<\yngLab[1pt]{2s-3}>,;}\,,
\end{equation}
or explicitly
\begin{equation}
    \omega_{A(2s-2)} = \Sigma_C{}^B\,
    \big(\Phi_{A(2s-2)B}{}^C
    + \Theta_{A(2s-2),B}{}^C\big)
    + \Sigma_A{}^B\,\big(\zeta_{A(2s-3)B}
    + \Xi_{A(2s-3),B}\big)\,,
\end{equation}
where $\Phi$ is the gauge field we would like to keep,
and $\Theta$, $\zeta$ and $\Xi$ are $0$-forms parameterising
the aforementioned unwanted components. As it turns out,
the latter can be gauged away thanks to the algebraic/shift
symmetry generated by $\eta$. Indeed, decomposing 
this gauge parameter into its irreducible components as well, 
one finds that it contains the irreps
\begin{equation}
    \yngSU{;,;} \otimes \yngSU{:_4<\yngLab[1pt]{2s-3}>,!\antiFund;} \quad\cong\quad \yngSU{:_5<\yngLab[1pt]{2s-2}>,:;,!\antiFund;}\
    \oplus\ \yngSU{::_4<\yngLab[1pt]{2s-4}>,!\antiFund;;}\ \oplus\ \yngSU{_5<\yngLab[1pt]{2s-2}>}\ \oplus\ \yngSU{_4<\yngLab[1pt]{2s-3}>,;}
\end{equation}
or explicitly
\begin{equation}
    \begin{aligned}
        \eta_{A(2s-3)}{}^B & = e^{C,D}\,
        \big(\Theta_{A(2s-3)C,D}{}^B
        + \varepsilon_{\times CDA}\,
        \Lambda_{A(2s-4)}{}^{B\times} \\
        & \hspace{80pt} + \delta^B_C\,\zeta_{A(2s-3)D}
        + \delta^B_C\,\Xi_{A(2s-3),D}
        -\tfrac{2s-3}{2s}\,\delta^B_A\Xi_{A(2s-4)C,D}\big)\,,
    \end{aligned}
\end{equation}
where $\Theta$, $\zeta$ and $\Xi$ are \emph{not}
the same tensors as those appearing in the decomposition
of $\omega$, but carry the same irrep (which is why 
we choose to denote them by the same symbols).
They can therefore be used to gauge away the unwanted
components in $\omega$. The last tensor $\Lambda$
fortunately \emph{does not appear} in the gauge transformation
\eqref{eq:gauge_transfo}, as can be checked
using the identity \eqref{eq:product}. Finally,
the differential gauge parameter $\xi$ can be decomposed into,
\begin{equation}
    \yngSU{;,;} \otimes \yngSU{_4<\yngLab[1pt]{2s-2}>}
    \quad\cong\quad \yngSU{_5<\yngLab[1pt]{2s-1}>,;}\
    \oplus\ \yngSU{:_4<\yngLab[1pt]{2s-3}>,!\antiFund;}\,,
\end{equation}
i.e. it is parameterised by two tensors via
\begin{equation}
    \xi_{A(2s-2)} = e^{B,C}\,\xi_{A(2s-2)B,C}
    + e_{B,A}\,\lambda_{A(2s-3)}{}^B\,,
\end{equation}
where we abused notation by using the same symbol
for the component of $\xi_{A(2s-2)}$ which corresponds
to the gauge parameter involved in the gauge transformation
\eqref{eq:gaugePhi} of $\Phi_{A(2s-1)}{}^B$,
that is $\xi_{A(2s-1),B}$. One can check,
by direct computation, that the gauge transformation
generated by the extra component $\lambda$
\emph{does not} affect the $\Phi$ component of $\omega$,
but only the pure gauge components $\Theta, \zeta$ and $\Xi$.

The equations of motion resulting from the previous action
are given by
\begin{subequations}
    \label{eq:EOM}
    \begin{eqnarray}
        0 & \approx & \nabla\Psi^{A(2s-2)BB} \wedge H_{BB}\,,\\
        0 & \approx & \nabla\omega_{A(2s-2)} \wedge H_{AA}\,,
    \end{eqnarray}
\end{subequations}
which are equivalent to the equations \eqref{eq:EOM_Weyl}
and \eqref{eq:EOM_potential}. Let us start with the first one,
which reads
\begin{equation}
    0 \approx \nabla_{C,D} \Psi^{A(2s-2)BB}\,
    e^{C,D} \wedge H_{BB}
    = 2\,\Sigma_B{}^E \wedge \Sigma_E{}^C\,
    \nabla_{C,D} \Psi^{A(2s-2)BD}\,,
\end{equation}
upon using \eqref{eq:1times3form}, and hence
does reproduce \eqref{eq:EOM_Weyl}. We already saw
that the second equation is nothing but the equation
for $\Phi$ embedded in $\omega$, see \eqref{eq:potential_form},
and that all other components can be gauged away,
hence the pair of wave equation \eqref{eq:EOM}
is equivalent to the equations \eqref{eq:EOM_Weyl}
and \eqref{eq:EOM_potential}.

\paragraph{Reducibility of the gauge symmetries.}
Before going further, let us spend more time
on the free gauge symmetries \eqref{eq:gauge_transfo},
with a special attention to their \emph{reducible} character.
As discussed above, this last property is necessary
in order to ensure that the free action \eqref{eq:free_action}
and the wave equations that stem from it describes
the propagation of the correct number of degrees of freedom,
namely $2s+1$ per fields (i.e. for $\Psi$ and for $\omega$).
This corresponds to the dimension of the irrep of the $6d$ massless 
little group $Spin(4)$, with highest weight $(s,\pm s)$,
and can be checked using the algorithm spelled out
in \cite{Kaparulin:2012px}.%
\footnote{Let us also mention that the fact that both
$\Psi$ and $\Phi$ (and hence by extensions $\omega$)
propagate the same number of degrees of freedom follows
from the general analysis of \cite{Lyakhovich:2025sau}.
I am thankful to an anonymous referee for pointing out
these references, as well as insightful remarks about
the counting of degrees of freedom.}

We should start by remarking that the free description
outlined in the previous section can be understood
in terms of the Lie algebra $\so(2,5)$,
and its finite-dimensional representation labelled
by the three-row Young diagram of length $s-1$,
whose decomposition under the Lorentz subalgebra reads
\begin{equation}\label{eq:branching}
    \gyoung(_5<\yngLab[1pt]{s-1}>,_5,_5) \quad
    \overset{\so(2,5)}{\underset{\so(1,5)}{\downarrow}} \quad
    \bigoplus_{\sigma=1}^s\ \yngSO{_5<\yngLab[1pt]{s-1}>,%
    _5,_4<\yngLab[1pt]{s-\sigma}>}_+ 
    \oplus \yngSO{_5<\yngLab[1pt]{s-1}>,%
    _5,_4<\yngLab[1pt]{s-\sigma}>}_-
    \cong\ \bigoplus_{\sigma=0}^{2s-2}\
    \yngSU{:::_5<\yngLab[1pt]{2s-2-\sigma}>,%
    !\antiFund_3<\yngLab[1pt]{\sigma}>}
\end{equation}
where it should be understood that the two-row Young (sub)diagram
of $\so(1,5)$, with both rows of length $s-1$,
appears only once. To do so, one can start by parameterising
the space of solutions of the equations of motion
for the $2$-form $\omega$, as
\begin{equation}\label{eq:nabla_w}
    H_{AA} \wedge \nabla \omega_{A(2s-2)} \approx 0
    \qquad\Longrightarrow\qquad
    \nabla \omega_{A(2s-2)}
    + e_{A,B} \wedge \omega_{A(2s-3)}{}^B
    = H^{BB}\,C_{A(2s-2)BB}\,,
\end{equation}
where $\omega_{A(2s-3)}{}^B$ is a $2$-form
and $C_{A(2s)}$ a $0$-form. Having introduced 
these two new fields, which describe
the \emph{unconstrained} part of the first derivatives
of $\omega_{A(2s-2)}$, one should try and characterise them.
To do so, we can differentiate the above equation
and parameterise the first derivatives of the new fields
in a way consistent with the previous equation. 
For the $2$-form $\omega_{A(2s-3)}{}^B$, this leads to
\begin{equation}
    \nabla \omega_{A(2s-3)}{}^B
    + e_{A,C} \wedge \omega_{A(2s-4)}{}^{BC}
    + (2s-2)\,e^{B,C} \wedge \omega_{A(2s-3)C} = 0\,,
\end{equation}
where we have introduced a new $2$-form $\omega_{A(2s-4)}{}^{BB}$.
Its appearance is due to the freedom afforded by
the fact that the operation which consists in wedging
a form $\alpha_{A(2s-3)}{}^B$ with a vielbein
and contracting one of its index with $B$ and symmetrising
the other with the $A$'s, i.e. the operation
$\alpha_{A(2s-3)}{}^B
\longmapsto e_{A,B} \wedge \alpha_{A(2s-3)}{}^B$,
has a non-trivial kernel. The $3$-form
$e_{A,C} \wedge \omega_{A(2s-4)}{}^{BC}$ belongs to this kernel,
and hence is consistent with the equation \eqref{eq:nabla_w}
for $\nabla\omega_{A(2s-2)}$ in that it does not appear
when differentiating it.

The spectrum of $\so(1,5) \cong \su^*(4)$-valued $2$-forms
appearing in this process is exactly that
of the branching rule \eqref{eq:branching} above.
Moreover, these $2$-forms \emph{do assemble}
into the $\so(2,5)$-irrep labelled by the three-row diagram
of length $s-1$, as can be verified by analysing the terms
proportional to the vielbein $e_{A,B}$ in their equations
of motion. Indeed, it turns out that the latter stem
from the action of the `transvection' generators, 
i.e. the generators of the complement of the Lorentz
subalgebra in $\so(2,5)$. To see that, it is useful
to introduce auxiliary variables $y^A$ and $\bar y_A$
to contract the indices carried by the $2$-forms with, 
i.e. we define
\begin{equation}
    \omega^{(-t)} := \frac1{(2s-t-2)!\,t!}\,y^{A(2s-t-2)}
    \bar y_{B(t)}\,\omega_{A(2s-t-2)}{}^{B(t)}\,,
\end{equation}
for $t=0,1,\dots,2s-2$. The action of $\su^*(4)$
can be represented by 
\begin{equation}
    \rho\big(M^A{}_B\big) = y^A\,\tfrac{\partial}{\partial y^B}
    - \bar y_B\,\tfrac{\partial}{\partial \bar y_A}\,,
\end{equation}
while the transvection generators by
\begin{equation}
    \rho\big(P^{A,B}\big) = y^{[A}\,\bar\partial^{B]}
    + \tfrac12\,\varepsilon^{ABCD}\,\bar y_C\,\partial_D\,.
\end{equation}
Upon introducing the operators
\begin{equation}\label{eq:sigma}
    \sigma_+ := e_{A,B}\,y^A\,\bar\partial^B\,,
    \qquad 
    \sigma_- := e^{A,B}\,\bar y_A\,\partial_B\,,
    \qquad 
    \sigma := \sigma_+ + \sigma_-
    \equiv e_{A,B}\,\rho\big(P^{A,B}\big)\,,
\end{equation}
one can rewrite the equations of motion
for the $2$-forms $\omega^{(-t)}$ as
\begin{equation}
    \big(\nabla+\sigma\big)\omega^{(-t)} 
    = \delta_{t,0}\,H^{BB}\,C_{A(2s-2)BB}\,
    \tfrac1{(2s-2)!}\,y^{A(2s-2)}\,,
\end{equation}
The operators $\sigma_\pm$ satisfy
\begin{equation}
    (\sigma_\pm)^2 = 0\,,
    \qquad \{\sigma_+,\sigma_-\}
    = \Sigma_A{}^B\,\rho\big(M^A{}_B\big)\,,
    \qquad\Longrightarrow\qquad
    \big(\nabla + \sigma\big)^2 = 0\,,
\end{equation}
and relate $2$-forms $\omega^{(k\mp1)}$ to $\omega^{(k)}$,
i.e. they increase/decrease the upper index by one unit.
Since the equations of motion are given in terms
of the action of a differential, $\nabla+\sigma$,
on the $2$-forms, their gauge symmetry simply takes the form
of exact terms, 
\begin{equation}\label{eq:trivial_parameters}
    \delta_\xi\omega^{(-t)} = \nabla\xi^{(-t)}
    +\sigma_-\xi^{(-t+1)} + \sigma_+\xi^{(-t-1)}\,,
\end{equation}
where $\xi^{(-k)}$ is a $1$-form taking values
in the same irrep as $\omega^{(-k)}$---a diagram
with $k$ coloured boxes. In particular, for $t=0$,
two parameters enter the gauge transformation
of $\omega^{(0)}$ namely $\xi^{(0)}$ and $\xi^{(-1)}$,
which correspond to $\xi$ and $\eta$
as in the expression \eqref{eq:gauge_transfo} given above.
Similarly, the space of reducible gauge parameters
is simply that of exact ones, i.e. parameters of the form
\begin{equation}
    \mathring\xi^{(-t)} = \nabla\zeta^{(-t)}
    + \sigma_-\zeta^{(-t+1)} + \sigma_+\zeta^{(-t-1)}\,,
\end{equation}
where $\zeta^{(-k)}$ are now $0$-forms taking values
in the same irrep as $\xi^{(-k)}$. For $t=0$, we find that
two parameters $\zeta^{(0)}$ and $\zeta^{(-1)}$
enter the reducibility of $\xi^{(0)}$,
while for $t=-1$, these two parameters as well as
a third one, $\zeta^{(-2)}$, are involved
in the reducibility of $\xi^{(-1)}$. In this case,
$\zeta^{(0)}$ corresponds to the reducibility parameter
$\zeta$ used for the reduction \eqref{eq:reducibility}
gauge symmetry generated by $\xi$ and $\eta$, 
and constitute a consistent truncation of the above
general formula \eqref{eq:trivial_parameters}.
The various fields, gauge and reducibility parameters
discussed are summarised below, together with (some of)
the operators relating them.
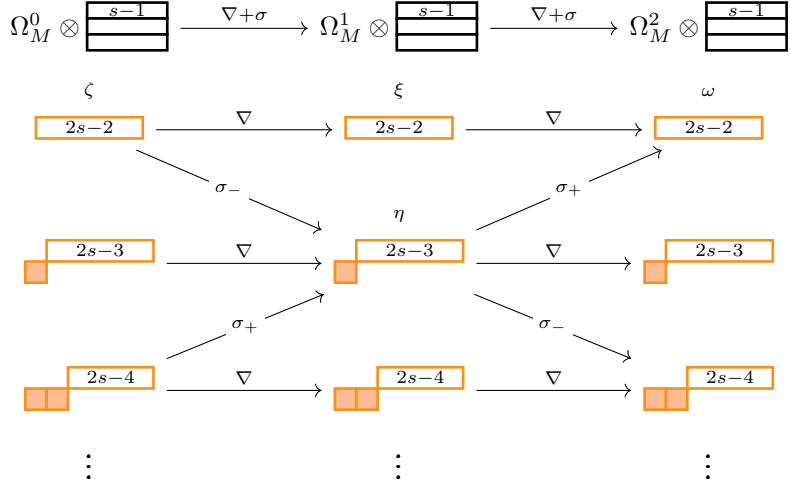
\begin{equation*}
    \begin{tikzcd}[column sep=huge]
        \Omega_M^0 \otimes {\Yboxdim{6pt}
        \gyoung(_5<\yngLab[-2pt]{s-1}>,_5,_5)}
        \ar[r, "\nabla+\sigma"]
        & \Omega_M^1 \otimes {\Yboxdim{6pt}
        \gyoung(_5<\yngLab[-2pt]{s-1}>,_5,_5)}
        \ar[r, "\nabla+\sigma"]
        & \Omega_M^2 \otimes {\Yboxdim{6pt}
        \gyoung(_5<\yngLab[-2pt]{s-1}>,_5,_5)} \\[-15pt]
        \overset{\zeta}{\yngSU{_5<\yngLab[-2pt]{2s-2}>}}
        \ar[r, "\nabla"]
        \ar[dr, crossing over, "\sigma_-" description]
        & \overset{\xi}{\yngSU{_5<\yngLab[-2pt]{2s-2}>}}
        \ar[r, "\nabla"] 
        & \overset{\omega}{\yngSU{_5<\yngLab[-2pt]{2s-2}>}} \\
        \yngSU{:_5<\yngLab[-2pt]{2s-3}>,!\antiFund;}
        \ar[r, "\nabla"] 
        & \overset{\eta}{\yngSU{:_5<\yngLab[-2pt]{2s-3}>,!\antiFund;}}
        \ar[r, "\nabla"]
        \ar[dr, "\sigma_-" description]
        \ar[ur, "\sigma_+" description]
        & \yngSU{:_5<\yngLab[-2pt]{2s-3}>,!\antiFund;} \\
        \yngSU{::_4<\yngLab[-2pt]{2s-4}>,!\antiFund;;}
        \ar[r, "\nabla"]
        \ar[ur, "\sigma_+" description]
        & \yngSU{::_4<\yngLab[-2pt]{2s-4}>,!\antiFund;;}
        \ar[r, "\nabla"]
        & \yngSU{::_4<\yngLab[-2pt]{2s-4}>,!\antiFund;;} \\[-15pt]
        \pmb\vdots & \pmb\vdots & \pmb\vdots
    \end{tikzcd}
\end{equation*}
Note that we did not draw all diagonal arrows
in the above picture, for the sake of clarity.

Splitting the action of the transvection generators
into two operators, $\sigma_\pm$, is a standard
and quite efficient way of analysing the content
of equations of motion which amount to fixing
the covariant derivative of a $p$-form valued
in a finite-dimensional module of some (isometry)
Lie algebra (to be non-vanishing only in a specific subspace).
In particular, one finds that the cohomology
of the $\sigma_-$ operator allows one to identify
more precisely this field content
(see e.g. \cite{Lopatin:1987hz, Bekaert:2004qos},
and references therein).

\subsection{A word about interactions}
\label{sec:interactions}
Now let us proceed with constructing a couple of examples
of interacting deformation of the previous free action.
As a first example, we will try to leverage the resemblance
of this free description with the $4d$ treatment
of massless higher spin fields based on $\so(1,3)\cong\sl(2,\C)$
and construct an action which maybe thought as the $6d$
counterpart of the higher spin extension of self-dual Yang--Mills 
introduced in \cite{Ponomarev:2017nrr, Krasnov:2021nsq}.

\paragraph{Warming-up: coupling to a connection $1$-form.}
First, let us show that the higher spin singletons
discussed in the previous section
can be coupled to a spin-$1$ gauge field, 
that is a $\g$-valued gauge field
$\GaugeYM \in \Omega^1_M \otimes \g$.
The Lie algebra $\g$ (whose indices and generators
we have suppressed for clarity) is assumed, as usual,
to be equipped with an ad-invariant symmetric bilinear form 
denoted by $\LieMetric{-,-}$.
In other words, this bilinear form satisfies
\begin{equation}\label{eq:sym_form}
    \LieMetric{x,y} = \LieMetric{y,x}
    \qand
    \LieMetric{[x,y]_\g,z} = \LieMetric{x,[y,z]_\g}\,,
\end{equation}
for any elements $x, y, z \in \g$.
Promoting the field $\Psi$ and $\omega$ to be $\g$-valued,
and defining the covariant derivative
and associated field strength,
\begin{equation}
    \CovDerYM := \nabla + [\GaugeYM,-]_\g\,,
    \qquad 
    \FieldStrengthYM  := {\rm d}\GaugeYM
    + \tfrac12\,[\GaugeYM,\GaugeYM]_\g\,,
\end{equation}
allows us to write the following action,
\begin{equation}\label{eq:action_YM}
    S_{\rm\sst YM-like}[\Psi,\omega;\GaugeYM]
    = \int_M \LieMetric{\Psi^{A(2s)},
    H_{AA} \wedge \CovDerYM \omega_{A(2s-2)}}\,,
\end{equation}
which is clearly invariant under the usual 
gauge transformations of a connection $1$-form,
\begin{equation}\label{eq:gauge_YM}
    \delta_\epsilon \GaugeYM = \CovDerYM\epsilon\,,
    \qquad 
    \delta_\epsilon \omega_{A(2s-2)}
    = [\omega_{A(2s-2)}, \epsilon]_\g\,,
    \qquad 
    \delta_\epsilon \Psi^{A(2s)}
    = [\Psi^{A(2s)}, \epsilon]_\g\,,
\end{equation}
with $\epsilon \in \Omega^0_M \otimes \g$.
On top of that, it can be made invariant under
\begin{equation}\label{eq:gauge_omega}
    \delta_{\xi,\eta} \omega_{A(2s-2)}
    = \CovDerYM\xi_{A(2s-2)}
    + e_{A,B} \wedge \eta_{A(2s-3)}{}^B\,,
\end{equation}
where both $\xi$ and $\eta$ are $\g$-valued $1$-forms,
by adding a BF term
\begin{equation}\label{eq:action_BF}
    S_{\sst\rm BF}[\GaugeYM, \BYM]
    = \int_M \LieMetric{\BYM,\FieldStrengthYM}\,,
    \qquad
    \delta_\epsilon \BYM = [\BYM, \epsilon]_\g\,,
\end{equation}
where $\BYM \in \Omega^4_M \otimes \g$ transform
in the adjoint representation of $\g$. This ensures
that this BF term is invariant under the gauge transformations
generated by $\epsilon$. More importantly, the introduction
of the $4$-form $B$ allow us to render the sum of the actions
$S_{\rm\sst YM-like}$ \eqref{eq:action_YM} 
and $S_{\rm\sst BF}$ \eqref{eq:action_BF} gauge-invariant
under the transformations \eqref{eq:gauge_YM}
generated by the parameters $\xi$ and $\eta$,
by a suitable choice of transformation for $B$.
More precisely, one can check that the variation
of $S_{\rm\sst YM-like}[\Psi,\omega;\GaugeYM]$
under the transformations generated by $\xi$ and $\eta$
is compensated by that of $S_{\sst\rm BF}[\GaugeYM,\BYM]$,
if one assumes that $\GaugeYM$ is inert and
\begin{equation}
    \delta_{\xi,\eta} \BYM 
    = -\big[\Psi^{A(2s)}, \xi_{A(2s-2)}\big]_\g \wedge H_{AA}\,.
\end{equation}
Doing so simply requires using the identities \eqref{eq:3times2form},
and the ad-invariance of the symmetric bilinear
form \eqref{eq:sym_form} on $\g$.

Finally, let us remark that the gauge transformations
of $\omega$ are reducible \emph{on-shell}, 
as the choice of gauge parameters
\begin{equation}
    \mathring\xi_{A(2s-2)} = \CovDerYM\zeta_{A(2s-2)}\,,
    \qquad 
    \mathring\eta_{A(2s-3)}{}^B
    = (2s-2)\,e^{B,C}\,\zeta_{A(2s-3)C}\,,
\end{equation}
in the presence of the gauge field $\GaugeYM$,
leads to trivial gauge transformations, for $F \approx 0$.%
\footnote{Note that one can even make the gauge transformations
reducible \emph{off-shell}, by adding to \eqref{eq:gauge_omega}
the term
$\tfrac1{3\,(s-1)}\,\big[\FieldStrengthYM,
    \,e^\mu_{A,B}\eta_{\mu|A(2s-3)}{}^B\big]_\g$,
which will imply a similar modification
of the gauge transformation of $B$.}

\paragraph{Higher spin extension.}
Paralleling the construction of \cite{Krasnov:2021nsq}
(see also \cite{Basile:2022mif} for the partially-massless case),
let us introduce the generating functions
\begin{equation}
    \begin{aligned}
        \Omega^0_M \otimes \C[\bar y]^{\Z_2} \otimes \g \ni 
        \Psi(x | \bar y) & := \sum_{s=1}^\infty \tfrac1{(2s)!}\,
        \bar y_{A(2s)}\,\Psi^{A(2s)}(x)\,, \\
        \Omega^2_M \otimes \C[y]^{\Z_2} \otimes \g \ni
        \omega(x | y) & := \sum_{s=1}^\infty \tfrac1{(2s-2)!}\,
        y^{A(2s-2)}\,\omega_{A(2s-2)}(x)\,,
    \end{aligned}
\end{equation}
both of which are again assumed to take values
in a Lie algebra $\g$.
Let us also define $H := \tfrac12\,H_{AA}\,y^A y^A$,
with the help of which we can write the sum
of the previous action \eqref{eq:action_YM}
for all integer spin $s\geq1$ in one go as, 
\begin{equation}
    S[\Psi,\omega;A,B] = \int_M p \circ \LieMetric{\Psi,
    H \wedge \CovDerYM\omega} + \LieMetric{B,F}\,,
\end{equation}
where 
\begin{equation}\label{eq:pairing}
    \begin{aligned}
        p: \C[\bar y] \otimes \C[y]
        & \longrightarrow \qquad \C \\
        f(\bar y) \otimes g(y)
        & \longmapsto\
        \sum_{n=1}^\infty \tfrac1{n!}\,f^{A(n)} g_{A(n)}\,,
    \end{aligned}
\end{equation}
is the pairing between totally symmetric $\su^*(4)$-tensors
and their conjugate. The gauge parameters
can also be packaged into generating functions
\begin{equation}\label{eq:free_gauge}
    \xi = \sum_{s=1}^\infty \tfrac1{(2s-2)!}\,
    y^{A(2s-2)}\,\xi_{A(2s-2)}\,,
    \qquad 
    \eta = \sum_{s=2}^\infty \tfrac1{(2s-3)!}\,
    y^{A(2s-3)}\,\bar y_B\,\eta_{A(2s-3)}{}^B\,,
\end{equation}
both of which are $1$-form and valued in the Lie algebra $\g$,
so that the gauge symmetries of $\omega$ and $\Psi$
can be put into the compact form
\begin{equation}
    \delta_{\xi,\eta,\epsilon} \omega = \CovDerYM \xi
    + \sigma_+ \eta + [\omega,\epsilon]_\g\,,
    \qquad 
    \delta_\epsilon \Psi = [\Psi, \epsilon]_\g\,,
\end{equation}
where everywhere $[-,-]_\g$ should be understood
as the $\C[y,\bar y]$-linear extension of the Lie bracket of $\g$,
and the operator $\sigma_+$ is the operator defined previously%
---see \eqref{eq:sigma}. Similarly, the reducibility parameters
for these gauge symmetries can be simply written as
\begin{equation}
    \mathring\xi = \CovDerYM \zeta
    \qand
    \mathring\eta = \sigma_-\zeta\,,
\end{equation}
where $\zeta\in\Omega^0_M \otimes \C[y]^{\Z_2} \otimes \g$.
Finally, the gauge transformations of the field $\BYM$ read
\begin{equation}
    \delta_{\xi,\epsilon} \BYM 
    = -p\big([\Psi, \xi]_\g \wedge H\big) + [\BYM,\epsilon]_\g\,,
\end{equation}
where the pairing $p$ allowed us to package the contribution
required from all spin $s\geq1$. This also highlights
the fact that neither the gauge field $\GaugeYM$
nor the $\BYM$-field take values in the algebras
of (even) polynomials in $y^A$ or $\bar y_A$,
but this possibility is interesting to explore.

So let us now replace $\GaugeYM$ and $\BYM$
with $1$- and $4$- forms taking values
in $\C[y]^{\Z_2} \otimes \g$, i.e.
\begin{equation}
    \begin{aligned}
        \Omega^1_M \otimes \C[y]^{\Z_2} \otimes \g \ni
        \GaugeHSYM & := \sum_{s=1}^\infty \tfrac1{(2s-2)!}\,
        y^{A(2s-2)}\,A_{A(2s-2)}\,, \\
        \Omega^4_M \otimes \C[\bar y]^{\Z_2} \otimes \g \ni
        \BHSYM & := \sum_{s=1}^\infty \tfrac1{(2s-2)!}\,
        \bar y_{A(2s-2)}\,B^{A(2s-2)}\,,
    \end{aligned}
\end{equation}
and denote the covariant derivative
and curvature of $\GaugeHSYM$ as
\begin{equation}
    \CovDerHSYM := \nabla + [\GaugeHSYM,-]_\g\,,
    \qquad
    \FieldStrengthHSYM := \nabla\GaugeHSYM
    + \tfrac12\,[\GaugeHSYM,\GaugeHSYM]_\g\,,
\end{equation}
so that we can now consider the action
\begin{equation}\label{eq:action_hs}
    S[\Psi,\omega;\GaugeHSYM,\BHSYM]
    = \int_M p \circ \LieMetric{\Psi, H \wedge \CovDerHSYM\omega}
    + p\circ\LieMetric{\BHSYM, \FieldStrengthHSYM}\,.
\end{equation}
The $1$-form fields making up $\GaugeHSYM$,
though forming a topological background here
since they appear as part of a BF term,
can describe hook-symmetric $\su^*(4)$-tensors 
\begin{equation}
    \varphi_{A(2s-2)B,C}
    \quad\longleftrightarrow\quad
    \yngSU{_5<\yngLab[1pt]{2s-1}>,;}\ \cong\ 
    \yngSO{_5<\yngLab[1pt]{s}>,_4<\yngLab[1pt]{s-1}>,_4}_+\,.
\end{equation}
Indeed, such tensors are embedded in the $1$-form as
\begin{equation}
    A_{A(2s-2)} = e^{B,C}\,\varphi_{A(2s-2)B,C}
    + (\dots)\,,
\end{equation}
where the $(\dots)$ denote other irreps
that can be gauged away by a shift symmetry
(similar to the one considered for $\omega$).
If instead of a zero-curvature condition $\FieldStrengthHSYM=0$,
one considers the linear equations
$H \wedge \nabla\GaugeHSYM \approx 0$,
the solution space describes the irrep corresponding
to a massless mixed-symmetry field labelled by
the above Young diagram. 

Let us come back to the previous action
and postulate the same gauge transformations
for $\omega$ and $\GaugeHSYM$ as before,
except we replace the covariant derivative
$\CovDerYM \to \CovDerHSYM$, i.e.
\begin{equation}
    \delta_{\xi,\epsilon,\eta} \omega
    = \CovDerHSYM\xi + \sigma_+ \eta
    + [\omega,\epsilon]_\g\,,
    \qquad
    \delta_\epsilon \GaugeHSYM = \CovDerHSYM\epsilon\,,
\end{equation}
where 
\begin{equation}\label{eq:epsilon}
    \Omega^0_M \otimes \C[y]^{\Z_2} \otimes \g \ni
    \epsilon = \sum_{s=1}^\infty \tfrac1{(2s-2)!}\,
    y^{A(2s-2)}\,\epsilon_{A(2s-2)}\,,
\end{equation}
is also promoted to a generating function
for the gauge parameters associated
with the higher spin $\g$-valued fields making up $\GaugeHSYM$.
However, na\"ively adapting the gauge transformations
of $\Psi$ and $\BYM$ by replacing them,
as well as all gauge parameters, with the corresponding 
generating functions does not work: for instance,
$\delta_\epsilon\Psi = [\Psi,\epsilon]_\g$
does not belong to the same space as $\Psi$,
which is $\Omega^0_M \otimes \C[\bar y]^{\Z_2} \otimes \g$,
if $\epsilon \in \Omega^0_M \otimes \C[y]^{\Z_2} \otimes \g$
is the generating function \eqref{eq:epsilon}
\emph{and} the Lie bracket is understood
as the $\C[y, \bar y]$-linear extension of that of $\g$.

In order to gain insight into how to define
these gauge transformations, we can look at the variation
of the above action, which reads
\begin{equation}
    \begin{aligned}
    \delta_{\xi,\eta,\epsilon} S[\Psi,\omega;\GaugeHSYM,\BHSYM]
    & = \int_M p \circ \LieMetric{\delta_\epsilon\Psi,
                                H \wedge \CovDerHSYM\omega}
    + p \circ \LieMetric{\delta_{\xi,\epsilon}\BHSYM,
                                \FieldStrengthHSYM} \\
    & \hspace{50pt} - p \circ \LieMetric{\Psi, 
    \big([\FieldStrengthHSYM,\xi]_\g
    + [\CovDerHSYM\omega,\epsilon]_\g\big) \wedge H}\,.
    \end{aligned}
\end{equation}
As before, it seems natural to try and use the ad-invariance
of the form on $\g$ in order to compare the second line
with the first one, and thereby fix the gauge transformation
of $\Psi$ and $\BHSYM$. But to do so, we need to properly
extend this invariance property to forms
valued in $\C[y]\otimes\g$ or $\C[\bar y]\otimes\g$,
that is to formulate it in terms
of the $\C[y]$-linear extension of the Lie bracket of $\g$
and the bilinear form $p \circ \LieMetric{-,-}$. 
For this purpose, let us introduce the operation
\begin{equation}
    \bullet: \C[\bar y] \otimes \C[y]
    \longrightarrow \C[\bar y]
\end{equation}
defined by
\begin{equation}
    p\big(\psi \bullet f, g\big)
    = p\big(\psi, f \cdot g\big)
\end{equation}
or more concretely,
\begin{equation}\label{eq:bullet}
    (\psi \bullet f)(\bar y)
    = \sum_{m,n\in\N} \tfrac1{m!\,n!}\,
    \bar y_{A(m)}\,\psi^{A(m)B(n)} f_{B(n)}\,,
\end{equation}
for any $\psi \in \C[\bar y]$ and $f,g \in \C[y]$.
In plain words, $\bullet$ is a representation
of the commutative algebra $\C[y]$ on $\C[\bar y]$
seen as the dual\footnote{Note that the space $\C[\bar y]$
of polynomials in $\bar y$ can be identified
with the `restricted' dual of $\C[y]$.
By `restricted' dual, we simply aim at pointing out
the following subtlety: since the algebra $\C[y]$
is infinite-dimensional, one should be careful
when talking about its dual space. Having in mind
that this polynomial algebra is isomorphic
to the symmetric algebra of the vector irrep
of $\su^*(4)$, i.e. $\Yboxdim{5pt} \C[y] \cong S(\yngSU{;})$,
and that the algebra of polynomials in $\bar y$ 
is isomorphic to the symmetric algebra 
of its conjugate representation, the following inclusion holds
\[
    \Yboxdim{5pt} \C[\bar y] \cong S(\yngSU{!\antiFund;})
    \cong S(\yngSU{;}^*) \subset S(\yngSU{;})^* \cong \C[y]^*\,.
\]
In other words, the dual of the algebra of polynomials in $y$
\emph{contains} the algebra of polynomials in $\bar y$
as a subalgebra, which is merely an instance
of the standard fact that `the dual of an algebra of polynomials
is an algebra of formal power series'.}
of $\C[y]$ via the pairing $p$. We can now write
the identity
\begin{equation}\label{eq:invariance}
    p \circ \LieMetric{\psi, [f, g]_\g}
    = p \circ \LieMetric{\LieBullet{\psi}{f}, g}\,,
    \qquad 
    f,g \in \C[y] \otimes \g\,,\
    \psi \in \C[\bar y] \otimes \g\,,
\end{equation}
where we have introduced the notation $\LieBullet{-}{-}$
which should be understood as
\begin{equation}
    \LieBullet{(\psi \otimes X)}{(f \otimes Y)}
    := (\psi \bullet f) \otimes [X,Y]_\g\,,
    \qquad \psi \in \C[\bar y]\,,\ f \in \C[y]\,,
    \quad X, Y \in \g\,,
\end{equation}
i.e. the extension of the Lie bracket by $\C[y,\bar y]$-linearity,
composed with the bullet operation. We are now in position
of writing the gauge transformations of $\Psi$ and $\BHSYM$,
\begin{equation}\label{eq:gauge_HSBF}
    \delta_{\xi,\eta,\epsilon} \BHSYM
    = -\LieBullet{\Psi}{\xi \wedge H}
    + \LieBullet{\BHSYM}{\epsilon}\,,
    \qquad
    \delta_\epsilon \Psi = \LieBullet{\Psi}{\epsilon}\,,
\end{equation}
under which the action \eqref{eq:action_hs} is invariant,
thanks to the invariance property \eqref{eq:invariance}
of $p \circ \LieMetric{-,-}$.
Note also that the gauge symmetry of $\omega$
is still reducible (on-shell), with
\begin{equation}
    \mathring\xi = \CovDerHSYM\zeta
    \qand
    \mathring\eta = \sigma_-\zeta\,.
\end{equation}
In summary, one can couple the tower of higher spin
singletons described by the pair $(\omega,\Psi)$
to a `higher spin version' of a $\g$-valued connection $1$-form
upon introducing a `higher spin version' of a BF theory.

\paragraph{Adding a cubic term.}
Given a $3$-cocycle $\Cocycle \in \wedge^3 \g^*\otimes\g$
of the Lie algebra $\g$ valued in its adjoint representation,
one can also add another term to the action \eqref{eq:action_hs}, 
namely
\begin{equation}\label{eq:cubic}
    \int_M p \circ \LieMetric{\Psi,
    H \wedge \Cocycle(\GaugeHSYM,\GaugeHSYM,\GaugeHSYM)}\,,
\end{equation}
where $\Cocycle$ is linearly extended to $\C[y] \otimes \Omega_M$.
That $\Cocycle$ is a $3$-cocycle ensures that the combination
\begin{equation}
    \cG := \CovDerHSYM\omega
    - \tfrac16\,\Cocycle(\GaugeHSYM,\GaugeHSYM,\GaugeHSYM)\,,
\end{equation}
transforms as
\begin{equation}\label{eq:modified_gauge}
    \delta_\epsilon\cG = [\cG, \epsilon]_\g
    + \Cocycle(\FieldStrengthHSYM,\GaugeHSYM,\epsilon)\,,
\end{equation}
under the modified gauge transformations
\begin{equation}
    \delta_\epsilon\omega = [\omega,\epsilon]_\g
    + \tfrac12\,\Cocycle(\GaugeHSYM,\GaugeHSYM,\epsilon)\,.
\end{equation}
Assuming further that the bilinear form $\LieMetric{-,-}$
is \emph{non-degenerate}, one can modify the gauge transformation
of the field $\BHSYM$ as
\begin{equation}
    \delta_\epsilon\BHSYM = \LieBullet{\BHSYM}{\epsilon}
    + \widetilde\Cocycle\big(\Psi \bullet H,\GaugeHSYM,\epsilon)\,,
\end{equation}
where 
\begin{equation}
    \begin{aligned}
        \widetilde{\Cocycle}:\ & \g \otimes (\g\wedge\g)
        && \longrightarrow \qquad \g\\
        & x \otimes (y \wedge z) && \longmapsto\,
        \LieMetric{x,\Cocycle\big((-)^\#,y,z\big)}\,,
    \end{aligned}
\end{equation}
and $\#: \g^* \longrightarrow \g$ is defined by
$\LieMetric{\alpha^\#,x} \equiv \alpha(x)$
for any $\alpha\in\g^*$ and $x\in\g$. Explicitly,
if we choose a basis $\{\mathsf{e}_a\}$ of $\g$,
the components of $\widetilde{\Cocycle}$ read
\begin{equation}
    \widetilde{\Cocycle}_{abc}{}^d
    = \kappa_{ai}\,\Cocycle_{jbc}{}^i\,\kappa^{jd}\,,
\end{equation}
where $\kappa_{ab} := \LieMetric{\fe_a,\fe_b}$
are the components of the invariant bilinear form,
and $\kappa^{ab}$ those of its inverse
(i.e. $\kappa_{ac}\kappa^{bc}=\delta^b_a$).
This definition implies the identity
\begin{equation}
    \LieMetric{w,\Cocycle(x,y,z)}
    = \LieMetric{\widetilde{\Cocycle}(w,x,y),z}\,,
    \qquad \forall w,x,y,z\in\g\,,
\end{equation}
which in turn ensures that the action
\begin{equation}
    S[\Psi,\omega;\GaugeHSYM,\BHSYM]
    = \int_M p \circ \LieMetric{\Psi, H \wedge \cG}
    + p \circ \LieMetric{\BHSYM, \FieldStrengthHSYM}
\end{equation}
is invariant under the modified transformations 
\eqref{eq:modified_gauge} generated by $\epsilon$,
as well as the gauge transformations generated by $\xi$
and $\eta$ which are unaffected by the addition
of the cubic term \eqref{eq:cubic}.%
\footnote{Let us stress that the possibility of adding
such a cubic term is a specificity of the type of theory
under consideration here, that is, involving chiral fields
in $6$ dimensions.}

\paragraph{Non-linearities in the zero-forms.}
Finally, let us conclude this section by pointing out
that the free action \eqref{eq:free_action}
is of \emph{presymplectic AKSZ-type}
\cite{Alkalaev:2013hta, Grigoriev:2020xec},
meaning they are obtained by the AKSZ construction
\cite{Alexandrov:1995kv} with a presymplectic $\cQ$-manifold
as its target space instead of a symplectic one
as is usually assumed. This is reflected in the fact
\begin{equation}
    p\big(\Psi, H \wedge -): \Omega^3_M \otimes \C[y]
    \longrightarrow \Omega^6_M\,,
\end{equation}
has a kernel (as a consequence of $H_{AA} \wedge e_{A,B}=0$),
which ensures that the free action \eqref{eq:action_hs}
is gauge invariant. As a consequence,
the sum of the free actions for spin $s\geq1$ singletons
can be deformed to
\begin{equation}\label{eq:presymplectic}
    S[\Psi, \omega] = \int_M p\big(\Theta(\Psi),
    H \wedge \nabla\omega\big)\,,
    \where
    \Omega^0_M \otimes \C[\bar y] \ni
    \Theta(\Psi) = \Psi + \cO(\Psi^2)\,,
\end{equation}
is a $\C[\bar y]$-valued $0$-form which admits an expansion
in powers of $\Psi$ and starts with a linear term
(so as to recover the free theory). Note that
the presymplectic potential $\Theta(\Psi)$ for four-dimensional 
higher spin gravity was constructed up to second order
in the zero-form in \cite[Sec. 6]{Sharapov:2021drr},
building on the result of \cite{Sharapov:2016qne}
for the free theory.

On top of any polynomial in $\Psi$, one can construct
examples using the operator
\begin{equation}
    \Invariant_4 = \varepsilon_{ABCD}\,\varepsilon_{ABCD}\,
    \frac{\partial^2}{\partial\bar y_A\,\partial\bar y_A}
    \otimes\frac{\partial^2}{\partial\bar y_B\,\partial\bar y_B}
    \otimes\frac{\partial^2}{\partial\bar y_C\,\partial\bar y_C}
    \otimes\frac{\partial^2}{\partial\bar y_D\,\partial\bar y_D}\,,
\end{equation}
which acts on four $\Psi$ as
\begin{align}
    m_4 \circ \Invariant_4(\Psi^{\otimes 4})
    & = \varepsilon_{ABCD}\,\varepsilon_{ABCD}\,
    \sum_{s_k \in 2\N} \frac1{s_1!s_2!s_3!s_4!}\, \times\\
    & \nonumber \hspace{100pt} \times\
    \Psi^{AA\,E(s_1)}\,\Psi^{BB\,E(s_2)}\,
    \Psi^{CC\,E(s_3)}\,\Psi^{DD\,E(s_4)}\,
    \bar y_{E(s_1+s_2+s_3+s_4)}\,,
\end{align}
where we have also introduced the operator
$m_4: \C[\bar y]^{\otimes 4} \to \C[\bar y]$,
denoting the multiplication of (four) polynomials
of $\bar y_A$.\footnote{Note that if one only keeps
the spin $1$ sector, that is
$\Psi=\tfrac12\,\Psi^{AA}\,\bar y_A \bar y_A$, 
$m_4 \circ \Invariant_4$ is the invariant considered
in \cite[Sec. 4.3]{Avetisyan:2022zza}
for the democratic formulation of non-linear theories
for chiral $2$-forms.}
In other words, any $\Theta$ of the form
\begin{equation}
    \Theta(\Psi) = \Psi + \sum_{k\geq2} c_k \Psi^k
    + \sum_{l\geq1} \tilde c_l\,
    m_4 \circ \Invariant_4^k(\Psi^{\otimes 4})\,,
\end{equation}
for $c_k, \tilde c_k$ some (arbitrary real) coefficients,
can be used to define the interacting action
\eqref{eq:presymplectic}.

As a side comment, let us note that nonlinearities
in the zero-forms can also be introduced in the presence 
of a BF-system, provided one is given
a $(\C[y] \otimes \g)$-equivariant map
\begin{equation}
    \Theta: \bigoplus_{n\in\N}
        \big(\C[\bar y] \otimes \g\big)^{\otimes n}
    \longrightarrow \C[\bar y] \otimes \g\,,
\end{equation}
so that for $\Psi \in \Omega^0_M \otimes \C[\bar y] \otimes \g$
and transforming as in \eqref{eq:gauge_HSBF},
the polynomial $\Theta(\Psi)$ transforms similarly,
\begin{equation}
    \delta_\epsilon \Psi = \LieBullet{\Psi}{\epsilon}
    \qquad\Longrightarrow\qquad 
    \delta_\epsilon\Theta(\Psi)
        = \LieBullet{\Theta(\Psi)}{\epsilon}\,.
\end{equation}
In this case, we can replace $\Psi$ for $\Theta(\Psi)$
in the action \eqref{eq:action_hs} discussed above, i.e.
\begin{equation}
    S[\Psi,\omega;\GaugeHSYM,\BHSYM]
    = \int_M p \circ \LieMetric{\Psi, H \wedge \CovDerHSYM\omega}
    + p\circ\LieMetric{\BHSYM, \FieldStrengthHSYM}\,,
\end{equation}
which becomes gauge-invariant under
\begin{equation}
    \delta_{\xi,\epsilon,\eta} \omega
    = \CovDerHSYM\xi + \sigma_+ \eta
    + [\omega,\epsilon]_\g\,,
    \qquad
    \delta_\epsilon \GaugeHSYM = \CovDerHSYM\epsilon\,,
    \qquad 
    \delta_\epsilon \Psi = [\Psi,\epsilon]_\g
\end{equation}
provided the field $\BHSYM$ transforms as
\begin{equation}
    \delta_{\xi,\eta,\epsilon} \BHSYM
    = -\LieBullet{\Theta(\Psi)}{\xi \wedge H}
    + \LieBullet{\BHSYM}{\epsilon}\,,
\end{equation}
by the same mechanism as the one discussed
in the previous section.

Lastly, let us point out that one can also build
simple current-type interactions. To do so, suppose
that we have a \emph{$1$-form} $J^{A(2s)}$ which verifies
\begin{equation}\label{eq:conservation}
    \nabla J^{A(2s-2)BB} \wedge H_{BB} \approx 0\,,
\end{equation}
upon using the equations of motions (hence our use
of the symbol $\approx$ in the above equation).
One can then add the functional
\begin{equation}\label{eq:current_coupling}
    \int_M J^{A(2s)} \wedge H_{AA}
    \wedge \omega_{A(2s-2)}\,,
\end{equation}
to the free actions of all singletons involved,
i.e. that of spin-$s$, as well as all those involved
in the definition of $J^{A(2s)}$, and thereby obtain
a current-type interaction, which is \emph{on-shell}
gauge-invariant. Indeed, a simple computation shows
that such a term is invariant under the shift symmetry
of $\omega$ as a consequence of the property \eqref{eq:3times2form},
and invariant under the differential gauge symmetry
of $\omega$ \emph{on-shell}, thanks to the condition
\eqref{eq:conservation}. Such interaction terms could be
used as a starting point to define other interacting theories,
possibly involving additional fields not discussed here,
together with a number of singletons (potentially a finite one,
in the spirit of the current interactions discussed
in \cite{Basile:2022mif} or the more general analysis
of finite-spectrum higher spin theories in four dimensions
given in the recent \cite{Serrani:2025owx}).

Having a $1$-form conserved current as described
in the previous paragraph would constitute an example 
of a $1$-form symmetry, the simplest instance of higher-form
symmetries \cite{Gaiotto:2014kfa, Kapustin:2014gua}
(see e.g. \cite{Schafer-Nameki:2023jdn, Shao:2023gho}
for an introduction). Such symmetries are usually presented
in terms of a $2$-form current $\mathfrak{J}_{[2]} \in \Omega^2_M$,
which is conserved on-shell in the sense that
${\rm d}\!*\!\mathfrak{J}_{[2]} \approx 0$. This type of current
naturally couples to a $2$-form gauge field
$\mathfrak{O}_{[2]} \in \Omega^2_M$ via
$\int_M \mathfrak{O}_{[2]} \wedge *\mathfrak{J}_{[2]}$,
in \emph{any dimensions}. In our $6d$ example here,
the $2$-form current is the Hodge dual of the $4$-form 
$J^{A(2s-2)BB} \wedge H_{BB} \sim *\mathfrak{J}_{[2]}$,
so that the functional \eqref{eq:current_coupling}
is merely a re-writing of the expected current coupling term.

\section{Arbitrary even dimensions}
\label{sec:any_d}
The $d=4$ case \cite{Krasnov:2021nsq}
and the $d=6$ case discussed in the previous section
are so similar that it seems reasonable to expect
that the same approach can be successfully followed
in arbitrary \emph{even dimensions}, $d=2r$.
Taking a step back, one can notice that the field content
of the action \eqref{eq:free_action} in four
and six dimensions is captured by the following objects:
\begin{itemize}
\item A $0$-form
\begin{equation}
    \Psi_\pm^{a_1(s),a_2(s),\dots,a_r(s)}
    \qquad\longleftrightarrow\qquad
    \yngSO{_6<\yngLab[1pt]{s}>,'62,_6}_\pm
\end{equation}
taking values in the \emph{self-dual or anti-self-dual}
Lorentz irrep%
\footnote{Self-duality for such a $0$-forms means that
it verifies
\[
    \tfrac1{r!}\,\epsilon^{a_1 \dots a_r}{}_{b_1 \dots b_r}
    \Psi_\pm^{a_1(s-1)b_1, a_2(s-1)b_2,\dots,a_r(s-1)b_r}
    = \pm i^{r-2[\frac{r}2]}\,\tfrac1{r!}\,
    \Psi^{a_1(s),a_2(s),\dots,a_r(s)}\,,
\]
with $[\tfrac{r}2]$ the integer part of $\tfrac{r}2$, 
i.e. its `fiberwise Hodge dualisation' in \emph{any} column
is proportional to $\pm1$ (resp. $\pm i$)
times itself for $r$ even (resp. odd).}
(respectively denoted by a $+$ or $-$ subscript)
corresponding to a rectangular Young diagram of length $s$
and maximal height, that is $\tfrac{d}2=r$;
\item A $\tfrac{d-2}2=(r-1)$-form taking values in
\begin{equation}\label{eq:omega}
    \omega^{(-1)^r\pm}_{a_1(s-1),\dots,a_r(s-1)}
    \qquad\longleftrightarrow\qquad
    \yngSO{_5<\yngLab[1pt]{s-1}>,'52,_5}_{(-1)^r\pm}
\end{equation}
taking values in the (\emph{anti-)self-dual} Lorentz irrep
characterised by another rectangular Young diagram,
of (maximal) height $r$ and length $s-1$,
i.e. with one column less than the $0$-form.
\end{itemize}
For the sake of readability, we will hereafter suppress
the $\pm$ sub/superscript of $\Psi$ and $\omega$,
and will assume that $\Psi$ is self-dual
while $\omega$ is takes values in a self-dual irrep
if $r$ is even, and anti-self-dual if $r$ is odd.
We can then consider the action
\begin{equation}\label{eq:action_any_d}
    S[\Psi,\omega] = \int_M \Psi^{a_1(s), \dots, a_r(s)}\,
    e_{a_1} \wedge \dots \wedge e_{a_r}
    \wedge \nabla\omega_{a_1(s-1),\dots,a_r(s-1)}\,,
\end{equation}
where $\nabla$ is a connection of constant curvature,
\begin{equation}
    \nabla^2 = e^a \wedge e^b\,\rho(M_{a,b})\,,
\end{equation}
where $M_{a,b}$ are the generators of the Lorentz algebra
and $\rho$ the representation in which the fields
acted upon fall.\footnote{Here again we implicitly
normalise the curvature.} Note that the $r$-form
$e_{a_1} \wedge \dots \wedge e_{a_r}$ being contracted
with a self-dual tensor $\Psi$, only its self-dual part%
---if $r$ is even---or anti-self-dual part%
---if $r$ is odd---appears in the integrand.%
\footnote{Recall that in dimension $d=2r$,
the wedge product of two self-dual $r$-forms
vanish identically for $r$ odd and are proportional
to the volume form for $r$ even, while the wedge product
of a self-dual and an anti-self-dual $r$-form
identically vanishes for $r$ even and is proportional
to the volume form for $r$ odd.} 
The previous action is invariant
under the gauge transformations
\begin{equation}\label{eq:gauge_any_d}
    \begin{aligned}
        \delta_{\xi,\eta} \omega_{a_1(s-1),\dots,a_r(s-1)}
        & = \nabla\xi_{a_1(s-1),\dots,a_r(s-1)} \\
        & \hspace{50pt} + e_{a_r} \wedge
        \eta_{a_1(s-1),\dots,a_{r-1}(s-1),a_r(s-2)} 
        + (\dots)\,,
    \end{aligned}
\end{equation}
where the $(\dots)$ denote additional terms needed
in order to project the wedge product of a vielbein
with the $(r-2)$-form
\begin{equation}
    \eta_{a_1(s-1),\dots,a_{r-1}(s-1),a_r(s-2)}
    \qquad\longleftrightarrow\qquad
    \yngSO{_5<\yngLab[1pt]{s-1}>,%
    '52,_4<\yngLab[1pt]{s-2}>}\,,
\end{equation}
onto the symmetry of the rectangular diagram \eqref{eq:omega},
and make it traceless.

The \emph{algebraic} gauge symmetry generated by $\eta$
allows one to gauge away certain `unwanted' components
of the $\tfrac{d-2}2$-form $\omega$, which are represented by
the ellipsis in bracket below,
\begin{equation}
    r-1\,\left\{\,\parbox{10pt}{\yngSO{|4}}\right.
    \otimes\ \yngSO{_5<\yngLab[1pt]{s-1}>,'52,_5}
    \quad\cong\quad
    \yngSO{_6<\yngLab[1pt]{s}>,'62,_5<\yngLab[1pt]{s-1}>}\
    \oplus\ \pmb{\big[\cdots\big]}\,.
\end{equation}
From the point of view of the above tensor product,
the irreps comprised in $\pmb{\big[\cdots\big]}$
are \emph{traces}, and can be written as
\begin{equation} \label{eq:extra}
    \pmb{\big[\cdots\big]}\ \cong\ \bigoplus_{k=2}^r\
    (r-k)\,\left\{\,\parbox{10pt}{\yngSO{|3}}\right.
    \LR\ \overbrace{\yngSO{'52,'42,}}^{s-1}
\end{equation}
where $k-1$ boxes have been remove along the rightmost column
of the diagrams on the right of the symbol \parbox{12pt}{$\LR$}, 
which denotes the Littlewood--Richardson rule
for the tensor product of two Young diagrams (see e.g.
\cite[Sec. 4]{Bekaert:2006py} for a pedagogical introduction).
These diagrams are recovered among the irreducible components
of the $\tfrac{d-4}2$-form $\eta$, which can again be sorted
into two pieces, 
\begin{equation}
    r-2\,\left\{\,\parbox{10pt}{\yngSO{|3}}\right.
    \otimes\ \yngSO{_5<\yngLab[1pt]{s-1}>,%
    '52,_4<\yngLab[1pt]{s-2}>}
    \quad\cong\quad \pmb{\big[\cdots\big]}\ 
    \oplus\ \pmb{\big\{\cdots\big\}}\,,
\end{equation}
where the extra term $\pmb{\big\{\cdots\big\}}$
can be expressed in a similar manner as \eqref{eq:extra}.
Though we will not give the precise decomposition
of this term here, one can check that it is contained
in the tensor product 
\begin{equation}
    r-3\,\left\{\,\parbox{10pt}{\yngSO{|3}}\right.
    \otimes\ \yngSO{'52,_4<\yngLab[1pt]{s-2}>,_4}
\end{equation}
corresponding to the $\so(1,d-1)$-irrep content
of the reducibility parameter of the gauge symmetry
\eqref{eq:gauge_any_d} of $\omega$. As it turns out,
this reducibility parameter contains again additional terms
on top of those included
in $\pmb{\big\{\cdots\big\}}$,
which are themselves contained in a $(r-4)$-form valued
in the diagram whose $r-3$ first row have length $s-1$
and the last $3$ have length $s-2$ (i.e. the diagram
obtained by removing three boxes along the rightmost column
of the rectangular diagram of height $r$ and length $s-1$).
Indeed, the latter is nothing but the second stage
reducibility parameter of the gauge transformations
considered here. This process goes on: the gauge symmetry
admits several stages of reducibility, whose parameters
are obtained by reducing simultaneously the form degree
and the number of boxes in the rightmost column
of the Lorentz diagram by one at each stage.
The Lorentz irreps appearing in the decomposition
of a parameter at a given stage of reducibility 
always split into two disjoint sets, each one contained 
in the decomposition of the parameters at either
the previous or the next stage. This ensures that $\omega$
only propagates the diagram obtained by removing one box
in the lower right corner of the rectangular diagram
of height $r$ and length $s$.

The equations of motion resulting
from the variation of the above action read
\begin{equation}
    \P_s\big(H_{a_1,\dots,a_r}^{(-1)^r} \wedge 
    \nabla\omega_{a_1(s-1),\dots,a_r(s-1)}\big) \approx 0
    \qand
    \nabla\Psi^{a_1(s-1)b_1,\dots,a_r(s-1)b_r}\,
    H^{(-1)^r}_{b_1,\dots,b_r} \approx 0\,,
\end{equation}
where $\P_s$ denotes the projection onto the diagram
which is a $r \times s$ rectangle,
and $H^\pm_{a_1,\dots,a_r}$ denotes the self-dual
or anti-self-dual part of the wedge product
$e_{a_1} \wedge \dots \wedge e_{a_r}$ of $r$ vielbeins.
Writing the self-dual and anti-self-dual part
of $\nabla\omega$ as,
\begin{equation}
    \nabla\omega_{a_1(s-1),\dots,a_r(s-1)}
    = H^{b_1,\dots,b_r}_+\,
    C^{(-1)^r}_{a_1(s-1)b_1,\dots,a_r(s-1)b_r}
    + H^{b_1,\dots,b_r}_-\,
    C^{-(-1)^r}_{a_1(s-1)b_1,\dots,a_r(s-1)b_r}
\end{equation}
where the sign of the superscript on the $0$-forms $C$
denotes its self-dual or anti-self-dual character,
and using the fact that
\begin{equation}
    H_{a_1,\dots,a_r}^{\mp} \wedge
    H_{b_1,\dots,b_r}^{(-1)^r\pm} = 0\,,
\end{equation}
the first equation implies that $\nabla\omega$
only contains an anti-self-dual piece, i.e. on-shell
\begin{equation}
    \nabla\omega_{a_1(s-1),\dots,a_r(s-1)}
    \approx H^{b_1,\dots,b_r}_-\,
    C^{-(-1)^r}_{a_1(s-1)b_1,\dots,a_r(s-1)b_r}
\end{equation}
and is therefore determined by a $0$-form
which is anti-self-dual for $r$ even,
and self-dual for $r$ odd. In other words, 
it describes a singleton of spin $s$
and of negative/positive chirality for $r$ even/odd.
The second equation of motion, that of $\Psi$,
allows one to parameterise its first derivative, 
by terms lying in the kernel of the map
which consists in contracting with the $r$-form
$H_{a_1,\dots,a_r}$, and repeat this process
for the $0$-forms introduced by imposing Bianchi identities.
This introduces new $0$-forms, and repeating
these steps \emph{ad infinitum}, one obtains
a system of first order differential equations
for infinitely many $0$-forms,%
\footnote{This procedure is known as `unfolding'
in the context of higher spin theories.}
describing the free propagation of a singleton
of spin $s$ described by the field $\Psi$,
and hence of positive chirality.
As a consequence, we find that the action 
\eqref{eq:action_any_d} discussed in this section
describes a pair of singleton of spin $s$
and of \emph{opposite/same chirality} in $d=2r$ dimensions
for $r$ even/odd. This is consistent with the $4d$ case
treated in \cite{Krasnov:2021nsq} and the $6d$ case
discussed in the previous section.

Note that free actions for conformal fields
of arbitrary mixed-symmetries have been studied
in details \cite{Vasiliev:2009ck},
and for higher spin singletons in particular
\cite{Marnelius:2008er, Marnelius:2009uw},
however in a slightly different form than the ones
we consider here.

\subsection{Adding colour}
Now that we have a free action in arbitrary even dimensions
of the same form as \eqref{eq:free_action} in $6d$,
it becomes straightforward to extend the (minimal) coupling
to a $\g$-valued connection $1$-form
described in the previous section:
nothing in this construction relies crucially
on the fact that $M$ is six-dimensional. 
The simplicity of the free field description is due to
the fact that higher spin singletons carry a particular
irrep of $\so(1,d-1)$, namely that corresponding
to a \emph{rectangular} Young diagram, of \emph{maximal} height.
Moreover, the BF fields in \eqref{eq:action_YM}
\emph{do not} mix singletons of different spins,
but couple to them individually. In other words,
we can write the action
\begin{equation}\label{eq:any_d}
    S[\Psi,\omega;\GaugeYM,\BYM]
    = \int_M \LieMetric{\Psi^{a_1(s),\dots,a_r(s)},
    H_{a_1,\dots,a_r} \wedge
    \CovDerYM\omega_{a_1(s-1),\dots,a_r(s-1)}}
    + \LieMetric{\BYM,\FieldStrengthYM}\,,
\end{equation}
where
\begin{itemize}
\item We used the notation $H_{a_1,\dots,a_r}
:= e_{a_1} \wedge \dots \wedge e_{a_r}$
with indices suppressed;
\item The $0$-form $\Psi^{a_1(s), \dots, a_r(s)}$,
the $\tfrac{d-2}2$-form $\omega_{a_1(s-1), \dots, a_r(s-1)}$,
the $1$-form $\GaugeYM$ and the $(d-2)$-form $\BYM$
all take values in a Lie algebra $\g$;
\item We again denoted the covariant derivative
with respect to the $\g$-valued gauge field $\GaugeYM$
as $\CovDerYM := \nabla + [\GaugeYM,-]_\g$,
and its curvature as $\FieldStrengthYM
:= {\rm d}\GaugeYM + \tfrac12\,[\GaugeYM, \GaugeYM]_\g$.
\end{itemize}
The action \eqref{eq:any_d} is readily verified
to be invariant under 
\begin{equation}
    \begin{aligned}
        \delta_{\xi,\epsilon} \omega & = \CovDerYM\xi
        + [\omega, \epsilon]_\g\,,
        &&& \delta_\epsilon \Psi & = [\Psi, \epsilon]_\g\,,\\
        \delta_\epsilon \GaugeYM & = \CovDerYM\epsilon\,,
        &&& \delta_{\xi,\epsilon} \BYM
        & = \P_0\big([\Psi, H \wedge \xi]_\g\big)\,,
    \end{aligned}
\end{equation}
where $\xi$ is a $(r-1)$-form whose indices,
as well as those of $\omega$ and $\Psi$, have been suppressed
for simplicity, $\epsilon$ is a $\g$-valued $0$-form,
in the trivial representation of the `fiber' Lorentz algebra
(i.e. not carrying any Latin indices) and $\P_0$
denotes the projection onto the trivial Lorentz irrep.
In other words, the indices of the $0$-form $\Psi$
are completely contracted with those of $H$ and $\xi$.
On top of that, the action is also invariant
under the transformations
\begin{equation}
    \delta_\eta \omega = \sigma_+ \eta\,,
\end{equation}
where $\sigma_\pm$ are defined as the piece of the action
of the transvection generators $P_a$ which relates $\so(1,d-1)$
irreps making up the $\so(2,d-1)$ irrep labelled by
a $(r+1) \times (s-1)$ diagram by respectively
adding or substracting a box in their Young diagram.
Schematically, these operators act as follows
\begin{equation}
    \begin{tikzcd}[column sep=huge, row sep=tiny]
        & \yngSO{_7<\yngLab[-2pt]{s-1}>,'72,_5<\yngLab[-2pt]{s-k}>;} \\
        \yngSO{_7<\yngLab[-2pt]{s-1}>,'72,_5<\yngLab[-2pt]{s-k}>}
        \ar[ur, "\sigma_+" description]
        \ar[dr, "\sigma_-" description] & \\
        & \yngSO{_7<\yngLab[-2pt]{s-1}>,'72,_4<\yngLab[-2pt]{s-k-1}>;<\yngLab[-2pt]{\times}>}
    \end{tikzcd}
\end{equation}
where all diagrams here correspond to irreps of Lorentz algebra
$\so(1,d-1)$ and hence are of height $r$.
As before, the gauge transformations generated by $\eta$
are meant to ensure that $\omega$ propagetes the correct number
of degrees of freedom on-shell.

In principle, one could push further and construct
a counterpart of the higher spin extension
of the above theory in arbitrary even dimensions:
to reproduce the $6d$ case, all one needs is a commutative
algebra to tensor with the Lie algebra $\g$,
thereby replacing the algebra of (even) polynomials
in $y^A$ used above, $\C[y]^{\Z_2}$. As explained previously,
the latter decomposes into a direct sum of finite-dimensional
$\su^*(4)$-modules, namely
\begin{equation}
    \C[y]^{\Z_2}\ \cong\ \bigoplus_{s=1}^\infty\, 
    \yngSU{_5<\yngLab[1pt]{2s-2}>}\,,
\end{equation}
where each one-row diagram of $\su^*(4)$ corresponds
to a three-row rectangular diagram of $\so(1,5)$,
whose length is half of that of the $\su^*(4)$ one.
This suggests looking for an algebra in $d=2r$ dimensions
whose decomposition consists of all rectangular $r \times (s-1)$
Young diagrams of $\so(1,d-1)$ with $s \geq 1$,
\begin{equation}
    \hs_\pm^{\tt ab}\ \cong\ \bigoplus_{s=1}^\infty\, 
    \yngSO{_5<\yngLab[1pt]{s-1}>,'52,}_\pm\,,
\end{equation}
and consider it as a commutative associative algebra.
One could think doing so by defining the multiplication
as the \emph{Cartan product}\footnote{The Cartan product 
of two finite-dimensional highest weight representations
of a semisimple Lie algebra is simply the projection
onto the irrep, in their tensor product,
whose highest weight is the sum of their two highest weights.
It is also the irrep with the maximal dimension 
in the tensor product. In terms of the two Young diagrams
associated with the highest weights, it simply corresponds
to gluing/concatenating them side-by-side (i.e. the length
of $k$th row of the new diagram is the sum of the length
of the $k$th row of the two original ones).}
of irreps of $\so(1,d-1)$, which indeed defines an associative
and commutative algebra \cite{Eastwood:2005ii}.
Explicitly realising this algebra would however involve
the use of projection operators for these rectangular diagrams,
which quickly become technically difficult to write down
(see e.g. \cite{Bulgakova:2022bsk} for recent work
on building traceless projectors for arbitrary Young diagrams).

\subsection{Some generalities}
Recall the first order formulation of Yang--Mills theory,
\begin{equation}
    S[\LagMult, A] = \int_M \big\langle\LagMult,
    F -\tfrac12\,\mathtt{g}_{\rm\sst YM}%
    \!*\!\LagMult\big\rangle_\g\,,
\end{equation}
where $F$ is the field strength of the Yang--Mills field
$A \in \Omega^1_M \otimes \g$ for $\g$
a Lie algebra with $\langle-,-\rangle_\g$
its invariant bilinear form, $\mathtt{g}_{\sst\rm YM}$
the Yang--Mills coupling constant, $*$ denotes the Hodge dual
on the $d$-dimensional manifold $M$,
and $\LagMult \in \Omega^{d-2}_M \otimes \g$
is a Lagrange multiplier. Integrating out $\LagMult$,
one recovers the usual, second order, Yang--Mills action.
In $4d$, one has the possibility of `truncating' this action
by keeping only the (say) self-dual part of $\LagMult$
and neglecting the term quadratic in it (by taking
the weak coupling limit $\mathtt{g}_{\sst\rm YM}\to0$).
Upon writing $\LagMult=\Psi^{\alpha\alpha}\,H_{\alpha\alpha}$
with $\Psi^{\alpha\alpha} \in \Omega^0_M \otimes \g$
the components of the self-dual part of the field strength,
and $H_{\alpha\alpha}$ a basis of self-dual $2$-forms,
written in terms of $\sl(2,\C)$-irreps,
or two-component spinors, one ends up with the action
\begin{equation}
    S[\Psi, A] = \int_M \big\langle \Psi^{\alpha\alpha}, 
    H_{\alpha\alpha} \wedge F \big\rangle_\g\,,
\end{equation}
considered as the starting point to obtain an higher spin
extension of self-dual Yang--Mills \cite{Krasnov:2021nsq}.

Since we are interested in theories containing
at least a $2$-form gauge fields, we can turn
our attention to a class of higher gauge theories 
(see e.g. \cite{Baez:2003fs, Borsten:2024gox})
based on two-term $L_\infty$-algebra, or Lie $2$-algebra.
For simplicity, let us focus on a \emph{minimal}%
\footnote{A minimal $L_\infty$-algebra is one for which
the differential is trivial. Recall that a Lie $2$-algebra
is a $L_\infty$-algebra structure on a two-term complex
\parbox{30pt}{$V \overset{\partial}{\longrightarrow} \g$}
 concentrated in degrees $-1$ and $0$.
This last restriction on the grading
implies that the only non-trivial brackets
(on top of the differential) are of arity two and three.
When the differential is trivial, such a structure
amounts to the data summarised above, namely a Lie algebra
structure on $\g$, a module structure on $V$,
and a $V$-valued $3$-cocycle. Whenever the differential 
is non-trivial however, the $L_\infty$-algebra structure
does not consists in these usual Lie theoretic objects,
as for starter the Jacobiator of the binary bracket on $\g$
no longer vanishes but is equal to the composition
$\partial \circ \Cocycle$, so that $\g$ is not a Lie algebra.}
Lie $2$-algebra, which boils down to the data
of a(n ordinary) Lie algebra $\g$ in degree $0$,
a module $V$ over it in degree $-1$,
and a Chevalley--Eilenberg $3$-cocycle
$\Cocycle \in \wedge^3\g^* \otimes V$ for $\g$ valued in $V$.
The Lie bracket on $\g$ and the representation of $\g$ on $V$
assemble into a binary bracket (of degree $0$)
on the graded vector space $\g \oplus V$ (hence concentrated
in degrees $0$ and $-1$), while the cocycle defines
a ternary bracket (of degree $-1$). The gauge fields 
associated with such an algebraic structure consists of
a $\g$-valued $1$-form $A \in \Omega^1_M \otimes \g$
and a $V$-valued $2$-form $\omega \in \Omega^2_M \otimes V$.
The gauge transformations of these fields are given by
\begin{equation}
    \delta_{\epsilon,\xi} A = {\rm d}\epsilon + [A,\epsilon]_\g\,,
    \qquad 
    \delta_{\epsilon,\xi} \omega = {\rm d}\xi + \rho(A)\xi
    - \rho(\epsilon)\omega + \tfrac12\,\Cocycle(A,A,\epsilon)\,,
\end{equation}
for $\epsilon\in\Omega^0_M \otimes \g$
and $\xi\in\Omega^1_M \otimes V$. The curvature 
of the $1$-form gauge field $A$ is unchanged (meaning,
it takes the same form as in the Yang--Mills case),
while the curvature of the $2$-form is defined,
and transforms, as
\begin{equation}
    G = {\rm d}\omega + \rho(A)\omega - \tfrac16\,\Cocycle(A,A,A)\,,
    \qquad 
    \delta_{\epsilon,\xi} G = -\rho(\epsilon)G + \rho(F)\xi
    + \Cocycle(F,A,\epsilon)\,.
\end{equation}
The important difference with respect to the usual
Yang--Mills field strength is that the curvature
of the $2$-form gauge field $\omega$ \emph{does not} transform
into \emph{itself} under (either types of)
the gauge transformations (unless, say $F=0$).
As a consequence, the na\"ive guess for an action
for $\omega$ that would be 
\begin{equation}\label{eq:naive}
    S_{\sst\rm YM-like}[A,\omega]
    = \tfrac12\,\int_M \big\langle G, *G \big\rangle_V
\end{equation}
where $\langle-,-\rangle_V$ is a $\g$-invariant%
\footnote{Meaning that it obeys
$\langle \rho(x)v, w\rangle_V + \langle v, \rho(x)w \rangle_V=0$
for any $x \in \g$ and $v, w \in V$.}
symmetric bilinear form on $V$, is not gauge invariant.
Instead, its gauge variation takes the form
\begin{equation}\label{eq:var_naive}
    \delta_{\epsilon,\xi} S = \int_M \big\langle \rho(F)\xi
    + \Cocycle(F,A,\epsilon), *G\big\rangle_V\,,
\end{equation}
i.e. gauge invariance of the na\"ive guess \eqref{eq:naive}
is obstructed by terms proportional to the field strength $F$.
This suggests the addition of a BF term as a way 
of restoring gauge invariance
(see e.g. \cite[Sec. 8]{Kotov:2010wr}):
indeed, the variation of the functional
\begin{equation}
    S_{\sst\rm BF}[B, A] = \int_M \big\langle B, F \big\rangle\,,
\end{equation}
with $B \in \Omega^{d-2}_M \otimes \g^*$
a $(d-2)$-form valued in the \emph{dual} of $\g$,
compensate the variation \eqref{eq:var_naive}
of the na\"ive guess \eqref{eq:naive}, provided that
the field $B$ transform as
\begin{equation}
    \delta_{\epsilon,\xi} B = [B, \epsilon]_{\g}^*
    - \big\langle\rho(-)\xi
        + \Cocycle(-,A,\epsilon),*G\big\rangle_V\,,
\end{equation}
where $[-,-]_{\g}^*$ denotes the coadjoint action of $\g$
on its dual $\g^*$ and the dash $(-)$ in $\rho$ and $\Cocycle$
denotes an empty entry for elements of $\g$
(since the whole expression is valued in $\g^*$).

At this point, we can easily find a first order formulation
for the gauge invariant action obtained above, namely 
\begin{equation}
    S[A,\omega;\LagMult, B] = \int_M \big\langle \LagMult,
    G - \tfrac12 *\!\LagMult \big\rangle_V + \langle B, F \rangle\,,
\end{equation}
i.e. we have introduced an auxiliary field
$\LagMult \in \Omega^{d-3}_M \otimes V$ such that 
its integration will set it to be $*G$, the Hodge dual
of the curvature of the $2$-form $\omega$.
Now fixing the dimension to $d=6$,
we can split the $3$-forms into self-dual
and anti-self-dual ones, and as in the $4d$ case
previously discussed, keep \emph{only the self-dual part}
of $\LagMult$ which we write as $\Psi^{AA}\,H_{AA}$
with $H_{AA}$ the basis of self-dual $3$-forms
in $\su^*(4)$ notation introduced in Section \ref{sec:free}.
After doing so (and neglecting the term quadratic in $\LagMult$),
one ends up with the action
\begin{equation}
    S[A,\Psi;\omega,B] = \int_M \big\langle \Psi^{AA},
    H_{AA} \wedge G \big\rangle_V + \langle B, F \rangle\,,
\end{equation}
which constitute the starting point of the higher spin
extension proposed here.

\section{Discussion}
\label{sec:discu}
Six-dimensional spacetime is known to offer
interesting possibilities for `exotic'---in terms
of field content---theories, including mixed-symmetry fields.
In this note, we discussed the case of higher spin singletons,
which are examples of such mixed-symmetry fields having
some very peculiar properties---for one thing they are conformal.
Thanks to the simple formulation of these fields afforded by
the isomorphism $\so(1,5) \cong \sl(2,\H) \cong \su^*(4)$,
we were able to derive a couple of examples
of interacting theories (which however break
conformal invariance) without encountering real difficulties
or subtleties. One can expect that other examples
should be possible to construct. For instance, 
considering that the field content involve $2$-forms,
one may expect the existence of higher-form currents
which could serve as the basis for building current interactions
as mentioned in the end of Section \ref{sec:interactions},
and thereby provide new examples of higher-form symmetries 
\cite{Gaiotto:2014kfa, Kapustin:2014gua}
(see e.g. \cite{Schafer-Nameki:2023jdn, Shao:2023gho}
for recent reviews).

Let us conclude this note by listing, in no particular order,
a few interesting directions and questions, that we would hope
to explore next.
\begin{itemize}
\item Considering that the contraction
of chiral higher spin gravity that is the higher spin
extension of self-dual Yang--Mills
seems to find some simple counterpart in $6d$,
as presented in this note, it is worth asking
whether or not the same is also true of the full theory.
In other words, does a counterpart
of chiral higher spin gravity exist in $6d$? 
If so, it would not only be interesting as another example
of complete higher spin theory, but also because
it would involve mixed-symmetry / higher forms gauge fields,
whose interactions are notoriously difficult to construct,
and scarce in examples.

\item As stated in the introduction, chiral higher spin gravity
in $4d$ admits a contraction to a higher spin extension
of self-dual gravity. Added to the fact that the singleton
of spin $s=2$ may be used for an exotic formulation
of gravity, one may expect that a $6d$ counterpart also exists.
It was recently proposed that the $4d$ higher spin
extension of self-dual gravity can be obtained
from a theory in $6d$, invariant under diffeomorphisms
for the total six-dimensional space \cite{Neiman:2024vit}.
One could speculate about the possibility of finding
a counterpart of this result, namely deriving
the putative $6d$ version of higher spin self-dual gravity
just mentioned from \emph{ten dimensions},
by extending spacetime with the four-dimensional spinor space
(for which the additional variables $y^A$ introduced to define
generating functions would be coordinates).

\item Another intriguing possibility would be
to obtain an higher spin version of the Penrose--Ward
correspondence in $6d$, in a similar manner as the one
recently derived in $4d$
\cite{Herfray:2022prf, Adamo:2022lah}.
As discussed in \cite[Sec. 5]{Mason:2011nw},
the related problem of defining a `twistor transform'
which establishes a bijection between solutions 
of the massless free equation for a spin $s$ singleton
in $6d$ and some cohomology class on twistor space
is not devoid of difficulties and subtleties
(see also \cite{Saemann:2011nb, Ivanova:2013vya}
and references therein). A closely related issue
would be to derive an action for theories
considered here in twistor space, following the works
\cite{Tran:2021ukl, Tran:2022tft, Adamo:2022lah}.
\end{itemize}

\section*{Acknowledgements}
I am grateful to Evgeny Skvortsov for numerous discussions
during the completion of this work, and to Victor Lekeu
for enlightening discussions on chiral $p$-forms,
and I am indebted to both of them for valuable feedback
on a previous version of this paper.
I am also thankful to Thanasis Chatzistavrakidis
and Sylvain Lavau for discussions on chiral $p$-forms,
tensor hierarchies and topics related to the content of this note.
Finally, I am grateful to the anonymous Referees
for their careful reading and insightful suggestions,
which have contributed to improving my understanding
of the subject as well as the quality of this manuscript.
This work was supported by the European Union’s
Horizon 2020 research and innovation programme
under the Marie Sk\l{}odowska Curie grant agreement No 101034383,
as well as from the European Research Council (ERC)
under the European Union’s Horizon 2020 research
and innovation programme grant agreement No 10100255.

\appendix
\section{Partially-massless extension}
\label{app:PM}
The higher spin extension of self-dual Yang--Mills
in $4d$ \cite{Krasnov:2021nsq} admits a `partially-massless'
counterpart \cite{Basile:2022mif}, that is a theory
whose spectrum consists of fields of arbitrary spin
and depth, valued in a Lie algebra $\g$. Recall that
gauge fields with higher derivative gauge transformations
are called `partially-massless' due to the fact
that they propagate an intermediary number
of degrees of freedom between massless and massive fields
\cite{Deser:1983mm, Higuchi:1986wu, Deser:2001us}.
The number of derivatives appearing
in their gauge transformations is usually called
the \emph{depth} and denoted $t$,
and hence the massless case corresponds to $t=1$.
Partially-massless fields of arbitrary mixed-symmetry 
have been analysed thoroughly in \cite{Boulanger:2008kw, Boulanger:2008up, Skvortsov:2009nv, Skvortsov:2009zu, Alkalaev:2009vm, Alkalaev:2011zv},
and the particular class considered here, with rectangular
Young diagrams, have been discussed in relations
with `higher-order' higher spin singletons \cite{Basile:2017kaz}.

Consider the action
\begin{equation}
    S[\Psi,\omega] = \int_M \Psi^{A(2s-t+1)}{}_{B(t-1)}\,
    H_{AA} \wedge \nabla\omega_{A(2s-t-1)}{}^{B(t-1)}\,,
    \qquad 1 \leq t \leq s\,,
\end{equation}
where, as previously, $\Psi$ and $\omega$ are a $0$-form 
and a $2$-form respectively,
subject to the gauge transformations
\begin{align}
    \delta_{\xi,\eta} \omega_{A(2s-t-1)}{}^{B(t-1)}
    & = \nabla\xi_{A(2s-t-1)}{}^{B(t-1)}
    + e_{A,C} \wedge \eta_{A(2s-t-2)}{}^{B(t-1)C} \\
    & \quad + e^{B,C} \wedge \tilde\eta_{A(2s-t-1),C}{}^{B(t-2)}
    - \tfrac{2s-t-1}{2s}\,\delta_A^B\,
    e^{C,D} \wedge \tilde\eta_{A(2s-t-2)C,D}{}^{B(t-2)}\,,
    \nonumber
\end{align}
with $\xi$, $\eta$ and $\tilde\eta$ being $1$-forms,
under which the action is invariant (again thanks
to the identities \eqref{eq:1times3form}
and \eqref{eq:3times2form} on the products of vielbeins).
These gauge symmetries are also reducible,
as it can be verified that the following specific choice
of gauge parameters
\begin{subequations}
    \begin{align}
        \mathring\xi_{A(2s-t-1)}{}^{B(t-1)}
        & = t\,\nabla\zeta_{A(2s-t-1)}{}^{B(t-1)}\,,\\
        \mathring\eta_{A(2s-t-2)}{}^{B(t)}
        & = 2t\,(s-t)\,e^{B,C} \zeta_{A(2s-t-2)C}{}^{B(t-1)}\,,\\
        \mathring{\tilde\eta}_{A(2s-t-1),C}{}^{B(t-2)}
        & = 2s\,(t-1)\,
        \big(e_{C,D}\,\zeta_{A(2s-t-1)}{}^{B(t-2)D}
        - e_{A,D}\,\zeta_{A(2s-t-2)C}{}^{B(t-2)D}\big)\,,
    \end{align}
\end{subequations}
leads to trivial gauge transformations.
The pair $(\Psi,\omega)$ describes a partially-massless field
with the symmetry a two-row rectangular diagram,
as in the bulk of the paper, but whose curvature has the symmetry
\begin{equation}
    \yngSO{_6<\yngLab[1pt]{s}>,_6,_4<\yngLab[1pt]{s-t+1}>}_+\
    \cong\ \yngSU{:::_5<\yngLab[1pt]{2s-t+1}>,%
    !\antiFund_3<\yngLab[1pt]{t-1}>}
\end{equation}
i.e. it is obtained by taking $s-t+1$ derivatives
of the gauge field. The massless case discussed previously
corresponds to $t=1$, while for $t>1$ the curvature
is defined with \emph{less} derivatives of the gauge field,
which reflects the fact that the latter
is subject to gauge transformations with more derivatives
of the gauge parameter.

As in the massless discussed in Section \ref{sec:interactions},
we can package the partially-massless fields of any spin
and depth into a generating $2$-form,
\begin{equation}
    \omega = \sum_{1 \leq t \leq s} \tfrac1{(2s-t-1)!\,(t-1)!}\,
    y^{A(2s-t-1)}\,\bar y_{B(t-1)}\,\omega_{A(2s-t-1)}{}^{B(t-1)}\,,
\end{equation}
and similarly for the $0$-forms $\Psi$,
as well as the $1$-forms $\GaugeHSYM$
and the $4$-forms $\BHSYM$
all of which are assumed to $\g$-valued. Extending
the definition of the pairing \eqref{eq:pairing} to
\begin{equation}
    \begin{aligned}
        p: \C[y, \bar y] \otimes \C[y, \bar y]
        & \longrightarrow \qquad\C\\
        f(y,\bar y) \otimes g(y, \bar y)
        & \longmapsto \sum_{m,n\in\N} \tfrac1{m!n!}\,
        f_{A(m)}{}^{B(n)}\,g_{B(n)}{}^{A(m)}\,,
    \end{aligned}
\end{equation}
and the action \eqref{eq:bullet} to all polynomials
in $y$ and $\bar y$ via
\begin{equation}
    p\big(\psi, f \cdot g\big) = p\big(\psi \bullet f, g\big)\,,
    \qquad \forall\,\psi,f,g\in\C[y,\bar y]\,,
\end{equation}
one can write the action
\begin{equation}
    S[\Psi,\omega;\GaugeHSYM, \BHSYM]
    = \int_M p\circ\LieMetric{\Psi, H \wedge \CovDerHSYM\omega}
    + p\circ\LieMetric{\BHSYM, \FieldStrengthHSYM}\,,
\end{equation}
and verify that it is invariant under the gauge transformations
\begin{subequations}
    \begin{align}
        \delta_\epsilon\Psi & = \LieBullet{\Psi}{\epsilon}\,,
        & \delta_{\xi,\epsilon} \omega & = \CovDerHSYM\xi
        + [\omega,\epsilon]_\g + \sigma_+\eta
        + \tilde\sigma_+\tilde\eta\,,\\
        \delta_\epsilon \GaugeHSYM & = \CovDerHSYM\epsilon\,,
        & \delta_{\xi,\epsilon} \BHSYM
        & = \LieBullet{\Psi}{\xi \wedge H}
        + \LieBullet{\BHSYM}{\epsilon}\,,
    \end{align}
\end{subequations}
where again, all gauge parameters are understood
to be generating functions, $\sigma_+$
is as defined in \eqref{eq:sigma},
and $\tilde\sigma_+$ is defined implicitly
from the part of the free gauge transformations
involving $\tilde\eta$.

\newpage
\providecommand{\href}[2]{#2}\begingroup\raggedright\endgroup

\end{document}